\documentclass[a4paper,english,11pt]{article}
\pdfoutput=1
\usepackage{jheppub}

\usepackage{epsfig,multicol}
\usepackage{amssymb,amsmath,bm, mathrsfs}
 \usepackage{amsfonts}
\usepackage{amstext}
\usepackage{epsf}
\usepackage{graphicx}
\usepackage{longtable}
\usepackage{afterpage}
\usepackage{placeins}
\usepackage{color}
\usepackage{bbm}
\usepackage{afterpage}

\usepackage{slashed}
\usepackage{textcomp}
\usepackage{placeins}

\newcommand{\lsim}{
\mathrel{\hbox{\rlap{\hbox{\lower4pt\hbox{$\sim$}}}\hbox{$<$}}}}
\newcommand{\gsim}{
\mathrel{\hbox{\rlap{\hbox{\lower4pt\hbox{$\sim$}}}\hbox{$>$}}}}

\newcommand{\gev}{\, {\rm GeV}}

\newcommand{\Heff}{{\cal H}_\text{ eff}}

\allowdisplaybreaks[2]
\newcommand{\be}{\begin{equation}}
\newcommand{\ee}{\end{equation}}
\newcommand{\bea}{\begin{eqnarray}}
\newcommand{\eea}{\end{eqnarray}}
\newcommand{\nn}{\nonumber}
\newcommand{\bi}{\begin{itemize}}
\newcommand{\ei}{\end{itemize}}
\newcommand{\ord}{{\cal O}}

\def\EmissT{\,{}/ \hspace{-1.5ex}  E_{T}}

\definecolor{orange}{rgb}{1,0.4,0.1}

\DeclareMathOperator{\diag}{diag}

\title{Flavored dark matter beyond Minimal Flavor Violation}

\author[a]{Prateek Agrawal,}
\author[b]{Monika Blanke,}
\author[a]{Katrin Gemmler}

\affiliation[a]{Fermilab, P.O. Box 500, Batavia, IL 60510, USA}
\affiliation[b]{CERN Theory Division, CH-1211 Geneva 23, Switzerland}

\emailAdd{prateek@fnal.gov}
\emailAdd{monika.blanke@cern.ch}
\emailAdd{katrin@fnal.gov}

\abstract{ We study the interplay of flavor and dark matter
phenomenology for models of flavored dark matter interacting with
quarks.  We allow an arbitrary flavor structure in the coupling of dark
matter with quarks.  This coupling is assumed to be the only new
source of violation of the Standard Model flavor symmetry extended by
a $U(3)_\chi$ associated with the dark matter.  We call this ansatz
\emph {Dark Minimal Flavor Violation} (DMFV)
and highlight its various implications, including an unbroken discrete symmetry that can
stabilize the dark matter.  As an illustration we study a Dirac
fermionic dark matter $\chi$ which transforms as triplet under
$U(3)_\chi$, and is a singlet under the Standard Model. 
The dark matter couples to right-handed down-type
quarks via a colored scalar mediator $\phi$ with a coupling $\lambda$.
We identify a number of ``flavor-safe'' scenarios for the structure of
$\lambda$
which are beyond Minimal Flavor Violation. For dark matter and
collider phenomenology we focus on the well-motivated case of
$b$-flavored dark matter. The combined flavor and dark matter
constraints on the parameter
space of $\lambda$ turn out to be interesting intersections of the
individual ones.
LHC constraints on simplified models of squarks and
sbottoms can be adapted to our case, and monojet searches
can be relevant if the spectrum is compressed. 
}

\keywords{Flavor, Dark Matter, Beyond the Standard Model}

\preprint{\\CERN-PH-TH-2014-098\\ FERMILAB-PUB-14-141-T}

\begin{document}

\maketitle


\section{Introduction}
\label{sec:int}

Dark matter (DM) provides a strong connection between the two
phenomenologically rich arenas: particle astrophysics and beyond
Standard Model (SM) physics. While the existence of DM is part of the
standard model of cosmology, its particle physics origins are largely
unknown. The WIMP (weakly interacting massive particle) miracle
however provides a tantalizing hint that DM is associated with 
new physics (NP) at the weak scale, and such candidates should be
accessible to various ongoing experiments. Signals at these
experiments depend strongly on the nature of interactions of the DM
with SM fields, and are less sensitive to other details of the model.
This motivates the study of simplified models, which minimally extend
the SM to include couplings of DM particles with the SM. Each
simplified model can then capture the dark matter phenomenology of a
wide range of models.

Once we consider different classes of simplified models, one new
category of models arises in analogy with SM flavor:
flavored
DM~\cite{Kile:2011mn,Kamenik:2011nb,Batell:2011tc,Agrawal:2011ze,
Batell:2013zwa,Kile:2013ola,Lopez-Honorez:2013wla,Kumar:2013hfa,
Zhang:2012da}.
In this setup DM particles come in multiple copies, and 
have a non-trivial flavor structure in their couplings with quarks and
leptons.\footnote{An alternative scenario in which dark matter arises from a discrete $A_4$ symmetry in the neutrino sector has been considered in \cite{Hirsch:2010ru,Boucenna:2011tj}.}
This framework does show up in a very
specific way in supersymmetric models as sneutrino DM models
\cite{Ibanez:1983kw, Ellis:1983ew, Hagelin:1984wv, Goodman:1984dc,
Freese:1985qw, Falk:1994es,MarchRussell:2009aq}, but clearly there are
more general possibilities.

\begin{figure}[h!]
\centering
\includegraphics[width=0.35\textwidth]{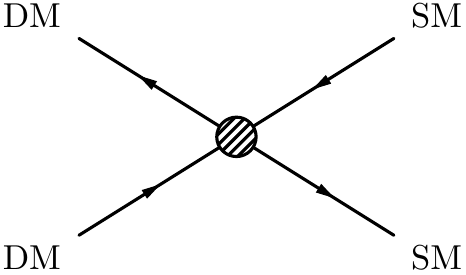}\qquad\qquad
\includegraphics[width=0.35\textwidth]{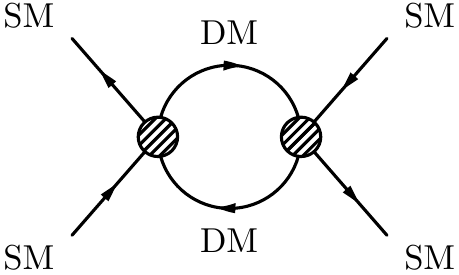}
\caption{Schematic diagrams contributing to experimental constraints on flavored DM.}
\label{fig:DMloop}
\end{figure}

This class of models is constrained, like other DM models, by both
indirect and direct detection DM experiments as well as collider
searches. The relevant schematic interaction responsible for these
signatures is shown in the left panel of figure \ref{fig:DMloop}.
Additionally precision flavor experiments have to be taken into
account due to the flavor violation introduced by the dark sector.
Schematically this contribution is displayed in the right panel of
figure \ref{fig:DMloop}, adding a new class of diagrams to the
well-studied DM-SM interaction.

Flavored dark matter models can have significantly distinct
phenomenology.  For indirect detection experiments, the spectrum of
photons and leptons arising from  DM annihilation depends on the
relative annihilation into various final states. For example, it was
shown that a DM candidate annihilating exclusively to $b$-quarks
provides a good fit to the spectrum of excess photons observed in a
recent analysis of Fermi-LAT data from the galactic
center~\cite{Daylan:2014rsa,
Berlin:2014tja,Agrawal:2014una,
Izaguirre:2014vva,
Boehm:2014hva,
Ipek:2014gua,
Kong:2014haa,
Ko:2014gha,
Boehm:2014bia,
Abdullah:2014lla,
Ghosh:2014pwa,
Martin:2014sxa,
Berlin:2014pya, 
Basak:2014sza,
Modak:2013jya} .  Direct detection predictions for scattering vary
widely depending upon whether the ambient DM particles couple to the
first generation quarks directly or not. The absence of direct
detection signals so far then point to the possibility of suppression
of such a coupling, which can be achieved through either a loop
suppression or a small mixing angle.
Collider searches for DM are also sensitive to the DM couplings to
various quark flavors, both in terms of the DM production cross section, 
as well as the flavor pattern of visible final states which 
can be produced in association with the DM. 

While some of these effects have been explored, the study of flavor
phenomenology has largely been restricted to elaborate models such as
the MSSM.  Previous analyses often assume for simplicity universality
or  minimal flavor violation (MFV)
\cite{Chivukula:1987py,Hall:1990ac,Buras:2000dm,D'Ambrosio:2002ex,Buras:2003jf},
so that flavor changing neutral current (FCNC) effects are
automatically suppressed.  On one hand this is welcome due to the
good agreement of the flavor data with the SM prediction, but on the
other hand interesting effects in the flavor sector are eliminated.

In this paper  we abandon the MFV principle and consider instead a
general flavor violating coupling of DM particles with quarks.  DM is
introduced as a triplet under a new global flavor symmetry
$U(3)_\chi$. While in our analysis the coupling matrix (denoted by
$\lambda$) is taken to be completely general, we make one simplifying
assumption that turns out to be helpful in various respects. We impose
that $\lambda$ is the only new source of flavor breaking, in addition
to the SM Yukawa couplings. As this assumption generalizes
the MFV principle to the DM sector, we call it {\it Dark Minimal
Flavor Violation} (DMFV). We will point out the following features of
DMFV:
\begin{itemize}
\item The DMFV framework, while bearing some conceptual similarity to MFV,
goes well beyond the latter framework, as it allows for large FCNC
effects. The structure of $\lambda$ needs to be determined from the
available constraints.
\item
  The DMFV ansatz naturally preserves a residual $\mathbb{Z}_3$
  symmetry, which guarantees the stability of the DM particle.
\item
DMFV significantly reduces the number of new parameters in the Lagrangian, as the DM mass term $m_\chi$ must be flavor conserving up to corrections of the form $\lambda^\dagger \lambda$.
\item
DMFV guarantees ``flavor-safety'' of the UV complete theory. It is therefore sufficient to identify flavor-safe scenarios for the structure of $\lambda$ within the simplified model framework.
\end{itemize}

In the phenomenological part of our paper we will restrict ourselves
to the study of the simplest version of DMFV, which we refer to as the
minimal DMFV (mDMFV) model in order to distinguish it from the more
general framework.  The DM is taken to be a Dirac fermion $\chi$, interacting
with the right-handed down-type quarks via the coupling
\begin{align}
\lambda \bar d_R \chi \phi
\end{align}
with a scalar mediator $\phi$.  While leaving the question of a
possible UV completion unanswered, this study captures the most
important phenomenological effects accessible to current experiments.
Our studies extend the existing literature on the phenomenology of
flavored DM in the following ways:
\begin{itemize}
 \item 
   {We go beyond the simple MFV hypothesis that automatically suppresses
 all flavor effects to an acceptable level. Instead we study the
 implications of a completely general coupling matrix $\lambda$,
 embedded in the DMFV ansatz, and derive its structure from the
 experimental constraints. }
 \item
 We consider a large number of relevant precision observables which can
 potentially be affected by the mDMFV model. These are in
 particular the constraints from meson-antimeson mixing, radiative and
 rare $B$ and $K$ decays, electroweak precision observables and
 electric dipole moments. 
\item
From the analysis of meson-antimeson mixing observables we identify a
number of ``flavor-safe'' scenarios for the structure of $\lambda$. These
scenarios will be useful for future studies of flavored DM
models beyond MFV, as they can be imposed simply and render detailed
re-analyses unnecessary.
\item
Subsequently we perform a simultaneous analysis of flavor and DM
constraints, such as the relic abundance from thermal
freeze-out, and direct detection data from LUX~\cite{Akerib:2013tjd}.
While restricting ourselves to the phenomenologically interesting case
of $b$-flavored DM, we consider several mass hierarchies in the
dark sector, i.\,e.\ large and small splittings between the DM
particle and the heavier flavors.
 \item 
We reveal a non-trivial interplay of the complementary flavor and DM
constraints, such that the combined constraint on the parameter
space of $\lambda$ turns out to be interesting intersections of the
individual ones.  This result underlines the importance of taking into
account the various constraints simultaneously.
 \item
We point out a cancellation between various mDMFV one-loop contributions
(photon penguin and box diagram) to the WIMP-nucleon scattering,
occurring for a certain range of coupling parameters. As the photon
penguin is only present for scattering off protons, while the box
diagram contributes to proton and neutron scattering cross-sections,
this cancellation provides a possible realization of Xenophobic
DM~\cite{Feng:2011vu,Feng:2013vod}.
\item 
We review the constraints from collider searches on the mDMFV model
with $b$-flavored DM. The most stringent bounds are placed by
searches for bottom squark pair production, constraining the parameter
space of the model up to a mediator mass $m_\phi\sim 800-900\gev$. Monojet
searches can be important for very compressed spectra, or for very
heavy $\phi$ such that its direct production is suppressed.
\end{itemize}

Our paper is organized as follows. In section \ref{sec:model} we
introduce the concept of Dark Minimal Flavor Violation (DMFV), and
describe the minimal model realizing this hypothesis, the mDMFV model.
Section \ref{sec:DMFV} deals with the implications of the DMFV
hypothesis that are valid beyond the minimal model. In section
\ref{sec:DF2constraints} we provide the formalism for a detailed study
of the constraints from meson anti-meson mixing on the mDMFV model. We
also consider potential new contributions to radiative and rare $B$
and $K$ decays, electroweak precision observables and electric dipole
moments and find all of these observables to be SM-like.
Section \ref{sec:DF2numerics} is devoted to a detailed numerical
analysis of the constraints on the coupling matrix $\lambda$ arising
from meson-antimeson mixing. We identify a number of ``flavor-safe'' scenarios for the coupling matrix $\lambda$. In section \ref{sec:flavor-consequences}
we provide a comprehensive summary of the results of the numerical
flavor analysis and the different scenarios emerging for the analysis
of DM constraints. In section \ref{sec:DMpheno} we study the
DM phenomenology of the mDMFV model, considering both the
relic abundance constraint from thermal freeze-out and the emerging
WIMP-nucleon cross section observed in direct detection experiments. A
combined numerical analysis of flavor and DM constraints is
performed in section \ref{sec:numerics}, studying the various possible
mass hierarchies in turn.  In section \ref{sec:collider} we estimate
the constraints on the mDMFV model from the LHC, stemming in
particular from monojet searches and searches for supersymmetric
bottom squarks. We also mention some new signatures for these models.
In section \ref{sec:conclusions} we summarize our results.  Some
technical details are relegated  to the appendices.


\section{Flavored dark matter beyond MFV -- a minimal model}\label{sec:model}
We consider a setup where DM $\chi$ transforms in the
fundamental representation of a new flavor symmetry $U(3)_\chi$, in
analogy with the SM flavor symmetry. While we posit this symmetry as
an ansatz, it will be an interesting future direction to study
possible UV completions (e.g. extended Grand Unified Theories) which
incorporate this structure.

We assume that the global
\be\label{eq:flavor-group}
U(3)_q \times U(3)_u \times U(3)_d \times U(3)_\chi
\ee
flavor symmetry is broken only by the SM Yukawa couplings $Y_u$,
$Y_d$ and the DM-quark coupling $\lambda$. This ansatz generalizes the
MFV hypothesis \cite{D'Ambrosio:2002ex,Buras:2000dm,Buras:2003jf,Chivukula:1987py,Hall:1990ac} to include an extra
$U(3)_\chi$ symmetry under which the DM field transforms, and an
additional Yukawa coupling $\lambda$. We refer to this
assumption as {\emph{Dark Minimal Flavor Violation (DMFV)}. 
Depending on the type of quark to which the DM couples, different
classes of DMFV can be defined, see appendix
\ref{app:DMFV-classification} for details. 

In what follows we restrict
ourselves to the coupling of $\chi$ to right-handed down-type quarks
via a scalar mediator
$\phi$. While the DM particle $\chi$ is a gauge singlet, the mediator
$\phi$ has to carry color and hypercharge. 
This helps to keep the model simple, since no further electroweak
structure is required when assuming the new particles to be singlets
under $SU(2)_L$. Further the choice of down-type quarks ensures to
have an effect in relevant flavor observables such as {$K$ and
$B_{d,s}$ meson mixing} and well-measured rare decays.

The most general renormalizable Lagrangian including the minimal field
content is then given by
\begin{eqnarray}
  {\cal L}&=& \mathcal{L}_\text{SM}+
  i \bar \chi \slashed{\partial} \chi
  - m_{\chi}  \bar \chi \chi  - (\lambda_{ij} \bar {d_{R}}_i \chi_j \phi + {\rm h.c.}) \nn \\
  && \qquad\,
  +
  (D_{\mu} \phi)^{\dagger} (D^{\mu} \phi) - m_{\phi}^2 \phi^{\dagger} \phi
  +\lambda_{H \phi}\, \phi^{\dagger} \phi\, H^{\dagger} H
  +\lambda_{\phi\phi}\, \phi^{\dagger} \phi\, \phi^{\dagger} \phi \,,\label{eq:Lagrangian}
\end{eqnarray}
with the symmetry transformation properties summarized in
table~\ref{tab:representations}. 
Note that the $U(3)_{\chi}$ flavor symmetry in the DM sector
guarantees that at the Lagrangian level all three DM flavors have the
same mass $m_{\chi}$, although they acquire a small splitting from
higher order DMFV corrections.  In what follows we refer to this model
as the {\emph{minimal DMFV (mDMFV) model}.

\begin{table}
  \begin{center}
    \begin{tabular}{|c||ccc|cccc|}
      \hline
      & $SU(3)_c$ & $SU(2)_L$ & $U(1)_Y$ & $U(3)_q$ & $U(3)_u$ & $U(3)_d$ & $U(3)_\chi$ \\ \hline\hline
      $q_L$ &  {\bf 3} & {\bf 2} & 1/6 & {\bf 3} & {\bf 1} & {\bf 1} & {\bf 1} \\
      $u_R$ & {\bf 3} & {\bf 1} & 2/3 & {\bf 1} & {\bf 3} & {\bf 1} & {\bf 1} \\ 
      $d_R$ & {\bf 3} & {\bf 1} & -1/3 & {\bf 1} & {\bf 1} & {\bf 3} & {\bf 1} \\ \hline
      $\ell_L$ &  {\bf 1} & {\bf 2} & -1/2 & {\bf 1} & {\bf 1} & {\bf 1} & {\bf 1} \\
      $e_R$ &  {\bf 1} & {\bf 1} & -1 & {\bf 1} & {\bf 1} & {\bf 1} & {\bf 1} \\ \hline
      $H$ &  {\bf 1} & {\bf 2} & 1/2 & {\bf 1} & {\bf 1} & {\bf 1} & {\bf 1} \\ \hline
      $\phi$ &  {\bf 3} & {\bf 1} & -1/3 & {\bf 1} & {\bf 1} & {\bf 1} & {\bf 1} \\ 
      $\chi_L$ &  {\bf 1} & {\bf 1} & 0 & {\bf 1} & {\bf 1} & {\bf 1} & {\bf 3} \\ 
      $\chi_R$ &  {\bf 1} & {\bf 1} & 0 & {\bf 1} & {\bf 1} & {\bf 1} & {\bf 3} \\ \hline\hline
      $Y_u$ &  {\bf 1} & {\bf 1} & 0 & {\bf 3} & \boldmath{${\bar 3}$} & {\bf 1} & {\bf 1} \\ 
      $Y_d$ &  {\bf 1} & {\bf 1} & 0 & {\bf 3} & {\bf 1} & \boldmath{${\bar 3}$}  & {\bf 1} \\ 
      $\lambda$  &  {\bf 1} & {\bf 1} & 0 & {\bf 1}  & {\bf 1} & {\bf 3} & \boldmath{${\bar 3}$}  \\ \hline
    \end{tabular}
  \end{center}
  \caption{Symmetry transformation properties of the minimal DMFV matter content and the Yukawa spurions.\label{tab:representations}}
\end{table}

The mDMFV model has some similarities to simplified models of supersymmetry and should be understood in an analogous manner. In contrast to the SUSY case however in mDMFV the flavor charge is carried by the DM fermions and not by the scalar mediator. Further we assume $\chi$ to be a Dirac fermion (a Majorana mass term would violate the $U(3)_\chi$ symmetry), while in the minimal SUSY models the gauginos are Majorana.

We stress that the DMFV ansatz, in contrast to the MFV ansatz,
potentially allows for large flavor violating effects. A careful
analysis of FCNC constraints is therefore necessary.

\section{Implications of the Dark Minimal Flavor Violation hypothesis}\label{sec:DMFV}

In the present section we consider the consequences of the DMFV
ansatz. We stress that these implications go beyond the simple mDMFV
model introduced in section~\ref{sec:model} and hold in any scenario
with the same DMFV flavor symmetry breaking pattern.

\subsection[New flavor violating parameters and a convenient parametrization for $\lambda$]{\boldmath New flavor violating parameters and a convenient parametrization for $\lambda$}

In the DMFV setup the flavor symmetry in the quark sector is broken only by the SM Yukawa couplings $Y_u$, $Y_d$ and the DM-quark coupling~$\lambda$. In
a first step the SM flavor symmetry can be used to remove unphysical parameters from the SM Yukawas. They can be parametrized as usual in terms of the six quark masses and the CKM matrix, signaling the misalignment between $Y_u$ and $Y_d$.

In the second step we remove unphysical parameters from the coupling matrix $\lambda$.
Being an arbitrary complex matrix, it contains at first 9 real parameters and 9 complex phases. Some of them can be removed by making use of the DM flavor symmetry $U(3)_\chi$. 

We start by parametrizing $\lambda$ in terms of a singular value decomposition
\be\label{eq:sing-val}
\lambda= U_{\lambda} D_{\lambda} V_\lambda\,,
\ee
where $D_\lambda$ is a diagonal matrix with real and positive entries,
and $U_{\lambda}$ and $V_\lambda$ are unitary matrices. Note that
$U_{\lambda}$ and $V_\lambda$ are not uniquely defined, as $\lambda$
is invariant under the diagonal rephasing
\be
U'_{\lambda} = U_{\lambda}\diag(e^{i\theta_1},e^{i\theta_2},e^{i\theta_3})\,,\qquad 
V'_{\lambda} = \diag(e^{-i\theta_1},e^{-i\theta_2},e^{-i\theta_3})V_{\lambda}\,.
\ee
We use this freedom to reduce the number of phases in $U_{\lambda}$ to
three. Then $\lambda$ has 9 real parameters and 9 phases in the
parametrization \eqref{eq:sing-val}. 
We can now use the  $U(3)_\chi$ invariance to fully remove the unitary matrix $V_\lambda$. Consequently we are left with the matrix 
\be\label{eq:sing-val-red}
\lambda= U_{\lambda} D_{\lambda} \,.
\ee
It contains nine parameters: three non-negative elements of
$D_{\lambda}$, and three mixing angles and three CP violating phases
in $U_{\lambda}$. Note that the mixing angles are restricted to the
range $0\le\theta_{ij}^\lambda\le\pi/4$ in order to avoid a
double-counting of parameter space. 
This choice ensures that each DM flavor couples dominantly to the
quark of the same generation. For instance we can refer to $\chi_3$ as
$b$-flavored DM.

A convenient parametrization for $U_{\lambda}$ has been derived in \cite{Blanke:2006xr} in the context of the Littlest Higgs model with T-parity. It can be written as
\bea
\addtolength{\arraycolsep}{3pt}
U_\lambda &=& U_{23}^\lambda U_{13}^\lambda U_{12}^\lambda \nn\\
&=&
 \begin{pmatrix}
  1 & 0 & 0\\
  0 & c_{23}^\lambda & s_{23}^\lambda e^{- i\delta^\lambda_{23}}\\
  0 & -s_{23}^\lambda e^{i\delta^\lambda_{23}} & c_{23}^\lambda\\
\end{pmatrix}
\begin{pmatrix}
  c_{13}^\lambda & 0 & s_{13}^\lambda e^{- i\delta^\lambda_{13}}\\
  0 & 1 & 0\\
  -s_{13}^\lambda e^{ i\delta^\lambda_{13}} & 0 & c_{13}^\lambda\\
\end{pmatrix}
\begin{pmatrix}
  c_{12}^\lambda & s_{12}^\lambda e^{- i\delta^\lambda_{12}} & 0\\
  -s_{12}^\lambda e^{i\delta^\lambda_{12}} & c_{12}^\lambda & 0\\
  0 & 0 & 1\\
\end{pmatrix}\,,\qquad
\eea
where $c_{ij}^\lambda = \cos\theta_{ij}^\lambda$ and  $s_{ij}^\lambda = \sin\theta_{ij}^\lambda$.
Performing the product one obtains the expression
\be
\addtolength{\arraycolsep}{3pt}
U_\lambda= \begin{pmatrix}
  c_{12}^\lambda c_{13}^\lambda & s_{12}^\lambda c_{13}^\lambda e^{-i\delta^\lambda_{12}}& s_{13}^\lambda e^{-i\delta^\lambda_{13}}\\
  -s_{12}^\lambda c_{23}^\lambda e^{i\delta^\lambda_{12}}-c_{12}^\lambda s_{23}^\lambda s_{13}^\lambda e^{i(\delta^\lambda_{13}-\delta^\lambda_{23})} &
  c_{12}^\lambda c_{23}^\lambda-s_{12}^\lambda s_{23}^\lambda s_{13}^\lambda e^{i(\delta^\lambda_{13}-\delta^\lambda_{12}-\delta^\lambda_{23})} &
  s_{23}^\lambda c_{13}^\lambda e^{-i\delta^\lambda_{23}}\\
  s_{12}^\lambda s_{23}^\lambda e^{i(\delta^\lambda_{12}+\delta^\lambda_{23})}-c_{12}^\lambda c_{23}^\lambda s_{13}^\lambda e^{i\delta^\lambda_{13}} &
  -c_{12}^\lambda s_{23}^\lambda e^{i\delta^\lambda_{23}}-s_{12}^\lambda c_{23}^\lambda s_{13}^\lambda e^{i(\delta^\lambda_{13}-\delta^\lambda_{12})} &
  c_{23}^\lambda c_{13}^\lambda\\
\end{pmatrix}.\ 
\ee

Finally it turns out to be convenient to parametrize the diagonal matrix $D_\lambda$ as
\be\label{eq:Dlambda}
D_\lambda \equiv \diag(D_{\lambda,11},D_{\lambda,22},D_{\lambda,33})=
\lambda_0\cdot \mathbbm{1}+\diag(\lambda_1,\lambda_2,-(\lambda_1+\lambda_2))\,.
\ee
The first parametrization is useful for the analysis of DM and collider constraints. The second parametrization instead is better suited for the flavor analysis, since it quantifies the deviations from a flavor universal coupling.

\subsection{Non-DMFV contributions and dark matter stability}

The flavor structure of the SM is accidental --- there exist no other
gauge-invariant operators beyond the Yukawa terms at the
renormalizable level. It is then worth asking if the DMFV ansatz
can also arise naturally in an analogous way.
In this section we study how generic the DMFV ansatz is from a UV
point-of-view. We stress however that the goal of this work is merely
to study the novel phenomenology arising from this ansatz, and a
complete UV model is beyond the current scope. We merely study the
corrections to the ansatz to the extent that they can affect low
energy phenomenology, particularly DM decay.

Interestingly, in the exact DMFV limit, all operators inducing decay
of the $U(3)_\chi$ triplet $\chi$ are forbidden, even at the
non-renormalizable level.  In analogy to the stability of DM
in  the MFV case \cite{Batell:2011tc} it can straightforwardly be
shown -- see appendix \ref{app:Z3} for details -- that  the flavor
symmetry \eqref{eq:flavor-group} broken only by the Yukawa couplings
$Y_u$, $Y_d$ and $\lambda$, together with $SU(3)_\text{QCD}$ imply an
unbroken $\mathbbm{Z}_3$ symmetry.  It is then natural to impose this
$\mathbb{Z}_3$ symmetry as exact, under which only the new particles
$\chi_i$ and $\phi$ are charged, and transform as
\begin{align}
  \chi_i \to e^{2\pi i/3} \chi_i, \qquad
  \phi \to e^{-2\pi i/3} \phi
  \ .
\end{align}
This symmetry prevents the decay of any of
these states into SM particles only, and therefore renders the
lightest state stable.

We now estimate the size of non-DMFV effects that can arise.
We imagine a UV scale, $\Lambda$, above which the DM flavor
symmetry $U(3)_\chi$ is unbroken. While this scale could in principle
be associated with the SM flavor scale as well, for simplicity we assume that
the SM flavor structure is generated at a higher scale. 
Generically, we expect all operators allowed by symmetries to be
generated at the scale $\Lambda$. The most important
contributions at low energy arise from relevant and marginal
operators.

The relevant operator 
\begin{align}
  \mathcal{O}_m
  &=
  \delta m_{ij} \, \bar{\chi}_i \chi_j
\end{align}
is the leading operator that is generated. It maximally violates the
DMFV ansatz, while preserving the $\mathbb{Z}_3$. This operator can be
prevented from being generated at the scale $\Lambda$ if the mass of
the DM fermions is generated at a
lower scale, through a flavor-blind sector. Note the analogy to
flavor-blind SUSY breaking, which yields MFV. It is an open question
whether such a scenario can be achieved simply in this framework. We
assume henceforth that this operator is negligible.

At the marginal level, we generate the following two operators,
\begin{align}
  \mathcal{O}_L
  &= \bar{\ell}_L \chi  H + \mathrm{h.c.}
  \\
  \mathcal{O}_B
  &= \bar{q}_L (i\sigma^2){q}_L^\dagger \phi^\dagger
  +\mathrm{h.c.},
\end{align}
which are lepton and baryon number violating respectively. These are
prohibited by the discrete $\mathbb{Z}_3$ symmetry, however.

We see that additional discrete symmetries can be imposed in order to
prevent the DM from decaying, and preventing non-DMFV
contributions at the renormalizable level. Higher dimensional non-DMFV
operators can then only modify other aspects of phenomenology, which
are not as severely constrained. Therefore, for a reasonably high
scale $\Lambda$, these operators are not expected
to alter the phenomenology appreciably.

\subsection{Mass splitting in the dark sector}

As noted in section \ref{sec:model} the DMFV hypothesis ensures that to leading order in the coupling $\lambda$, the
masses for different DM particles are equal. There are three
potential sources for splittings. 

Firstly, there can be contributions to the mass matrix $m_\chi$ directly
violating DMFV. We assume that such contributions are absent. Secondly,
higher dimensional DMFV-violating contributions can still induce splittings, but
these are expected to be suppressed by the heavy scale where DMFV is broken.

An unavoidable contribution is through the renormalization group running, where a universal
mass at the high scale is renormalized by the presence of the DM coupling $\lambda$ at low scales.
Generically, it is also possible that there is a DMFV preserving
contribution $\propto \lambda^\dagger\lambda$ at tree level. If present, this would be the largest
contribution to the DM splittings. Of course, the pattern of
splittings generated by the running and by such threshold effects is
identical, since both cases are consistent with DMFV.

The splittings are given by
\begin{align}\label{eq:DMFVmass}
  m_{ij}
  &=
  m_{\chi} (\mathbbm{1}
  +\eta\, \lambda^\dagger\lambda+ \cdots
  )_{ij}
  =
  m_{\chi} (1  + \eta (D_{\lambda, ii} )^2 + \cdots
  )\delta_{ij}\,,
\end{align}
where summation is not implied in the last term. Here $\eta$ is a real coefficient whose value depends on the details of the model.
If the contribution to the mass matrix arises at tree level, then $\eta$ is expected to be an $\mathcal{O}(1)$ number. On the other hand, the contribution from running is schematically given by
\begin{align}
  \eta
  &\sim
  \frac{1}{16\pi^2} \log \left(\frac{m_\chi^2}{\Lambda^2} \right)\,,
\end{align}
where $\Lambda$ is the dark flavor scale noted above.

The DMFV expansion above is only valid if higher order corrections are
parametrically suppressed. In order to ensure convergence, in what
follows we will assume $|\eta (D_{\lambda,ii})^2| <0.3$.

\section{Constraints from flavor and precision  observables}\label{sec:DF2constraints}

In this section we study all relevant constraints from flavor
observables on the mDMFV model. We start the analysis of the
well-measured and strongly constraining observables from meson
anti-meson mixing, followed by relevant rare decays. Finally we take a
brief look at electroweak precision tests and electric dipole moments.
While we will find that $\Delta F =2$ processes significantly shape
the structure of a phenomenologically viable coupling matrix
$\lambda$, effects of other flavor observables are negligible.

We restrict ourselves to providing the formulae directly relevant for our study, a more detailed description of relevant techniques and necessary formulae for the study of $\Delta F =2$ processes reaching from effective Hamiltonian to flavor observables can be found for instance in \cite{Blanke:2011ry}. A recent comprehensive review can be found in \cite{Buras:2013ooa}.

\subsection{Constraints from meson anti-meson mixing}

\begin{figure}[h!]
\centering
\includegraphics[width=0.35\textwidth]{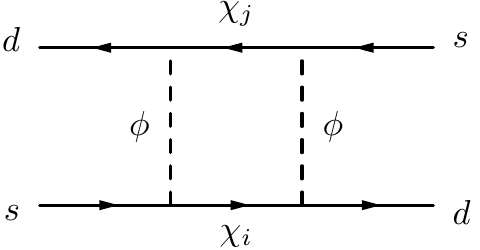}
\caption{New contribution to $K^0-\bar K^0$ mixing in the mDMFV model.}
\label{fig:kkbar}
\end{figure}

In the mDMFV model new contributions to $\Delta F =2 $ processes arise first at the one loop level. The relevant box diagram is shown in figure \ref{fig:kkbar} for the case of $K^0-\bar K^0$ mixing. Evaluating this diagram we obtain the following contribution to the effective Hamiltonian:
\be\label{eq:Heff1}
\Heff^{\Delta S=2,\text{new}} = \frac{1}{128\pi^2 m_\phi^2} \sum_{i,j}
\lambda_{si}\lambda_{di}^* \lambda_{sj}\lambda_{dj}^* F(x_i,x_j)
\times Q^{VRR} +\text{h.c.},
\ee
with $x_i=m_{\chi_i}^2/m_{\phi}^2$, and the loop function $F(x_i,x_j)$ can be found in appendix \ref{app:functions}. As the new particles $\phi$ and $\chi_i$ couple only to right-handed down-type quarks, the only effective operator which receives new contributions is
\be
Q^{VRR} = (\bar s_\alpha \gamma^\mu P_R d_\alpha) (\bar s_\beta \gamma_\mu P_R d_\beta)\,,
\ee
i.\,e. the chirality-flipped counterpart of the SM operator.

The mass splittings among the $\chi_i$ fields constitutes a
higher order correction in the DMFV expansion, which we assume to be
small. Thus we can take the limit of equal $\chi$ masses in
\eqref{eq:Heff1}. The effective Hamiltonian then simplifies to
\be\label{eq:Heff2}
\Heff^{\Delta S=2,\text{new}} = \frac{1}{128\pi^2 m_\phi^2} F(x) \, \xi_K^2 \times Q^{VRR} +\text{h.c.}\,,
\ee
where $x=m_\chi^2/m_\phi^2$. The loop function $F(x)$ can be found in appendix \ref{app:functions}. We also defined
\be
\xi_K = (\lambda\lambda^\dagger)_{sd} = \sum_{i=1}^3 \lambda_{si}\lambda_{di}^*\,.
\ee

The mDMFV contribution of the DM sector to the off-diagonal element of the $K^0-\bar K^0$ mass matrix can then be obtained from
\be
M_{12}^{K,\text{new}} = \frac{1}{2m_K}\langle \bar K^0| \Heff^{\Delta S=2,\text{new}} | K^0\rangle ^*\,.
\ee
Using
\be
\langle Q^{VRR}(\mu=2\,\text{GeV})\rangle =\frac{2}{3} m_K^2 F_K^2 \hat B_K
\ee
we obtain
\be
M_{12}^{K,\text{new}} = \frac{1}{384\pi^2 m_\phi^2} m_K F_K^2 \hat B_K \eta_2 F(x)  (\xi_K^*)^2 \,.
\ee
The parameter $\eta_2$ summarizes the corrections from the
renormalization group running from the weak scale $\mu\sim m_t$ down
to the scale $\mu=2\,$ GeV, where the lattice calculations are
performed, as well as the corrections due to the matching of the full
theory to the effective theory calculated within the SM.

By parametrizing the NLO corrections by $\eta_2$, we make two
approximations. We neglect the running from the NP scale
$\mu\sim m_\phi$ to the scale $\mu\sim m_t$, as well as the difference
in the matching conditions between the SM and the NP scenario studied
here.  In order to estimate the error associated to our approach, it is
useful to compare our case with the discussion of the 331 models in
\cite{Buras:2012dp}. In the latter framework the inclusion of the
next-to-leading order (NLO) corrections amounts to a few percent
correction to the size of the NP contribution. We expect similar
conclusions to hold also in our case, in particular since in the MSSM
the NLO corrections to the Wilson coefficient $C^{VRR}$ have been
found to be small \cite{Virto:2009wm}.

In an analogous manner we find
\be
M_{12}^{q,\text{new}} = \frac{1}{384\pi^2 m_\phi^2} m_{B_q} F_{B_q}^2 \hat B_{B_q} \eta_B F(x) (\xi_{B_q}^*)^2
\qquad (q=d,s)\,,
\ee
 where we define
\be\label{eq:xiBq}
\xi_{B_q} =  (\lambda\lambda^\dagger)_{bq} = \sum_{i=1}^3 \lambda_{bi}\lambda_{qi}^* \qquad (q=d,s)\,.
\ee

In passing we note that the mDMFV model, coupling only to down-type quarks, does not contribute to $D$ meson observables at the one loop level.

\subsection[Radiative and rare $K$ and $B$ decays]{\boldmath Radiative and rare $K$ and $B$ decays}

We now turn our attention to radiative and rare decays, starting with the electromagnetic dipole operator generating the $b\to s\gamma$ transition.

The effective Hamiltonian describing the $B\to X_s\gamma$ decay can be written as
\be
\Heff = \frac{4 G_F}{\sqrt{2}} V_{ts}^* V_{tb} \left( C_7 Q_7 + C'_7 Q'_7+\cdots \right)\,,
\ee
where we omitted the tree level and chromomagnetic dipole operators that contribute to $b\to s\gamma$ via renormalization group mixing. We use the normalization
\bea
Q_7 &=& \frac{e}{16\pi^2} m_b \bar s_L \sigma^{\mu\nu} b_R F_{\mu\nu}\,,\\
Q'_7 &=& \frac{e}{16\pi^2} m_b \bar s_R \sigma^{\mu\nu} b_L F_{\mu\nu}\,.
\eea
In the SM the Wilson coefficient $C'_7$ is strongly suppressed due to the chiral structure of weak interactions:
\be
C'_{7,\text{SM}} = \frac{m_s}{m_b} C_{7,\text{SM}}\,.
\ee

Conversely the DM in our scenario couples only to right-handed SM fermions -- therefore the only relevant new contribution arises in the chirality-flipped Wilson coefficient~$C'_7$. The relevant diagrams are analogous to the ones depicting the gluino contribution in supersymmetric models, replacing the gluino by the DM particles $\chi_i$ and the squarks by the scalar mediator $\phi$, and keeping only the coupling to right-handed SM quarks. Correcting for the different coupling and taking into account that $\chi_i$ are QCD singlets while the gluino is a color octet, we can straightforwardly obtain the result for $\delta C'_7$ from \cite{Bertolini:1990if,Cho:1996we}. Adjusting eq. (A.5) of \cite{Cho:1996we} to our model, we find
\be
\delta C'_7 =-\frac{1}{6 g_2^2 V^*_{ts} V_{tb}}\frac{m_W^2}{m_\phi^2} \, \xi_{B_s}^* \, g(x_i)\,,
\ee
where the short-hand notation for the relevant combination of elements of $\lambda$ has been defined in \eqref{eq:xiBq}, and the loop function $g(x)$ is given in appendix \ref{app:functions}. 

With $g(x) \sim 0.08-0.17$ the size of the new contribution $\delta C'_7$ can be estimated as 
\be
|\delta C'_7| \lsim 4\cdot 10^{-2} |\xi_{B_s}^*| \left[\frac{500\gev}{m_\phi}\right]^2 \,.
\ee
Comparing this result to the constraints on the size of NP
contributions, see e.\,g.\ figure~2 in \cite{Altmannshofer:2013foa}, we see that the effect on the electromagnetic dipole operators generated in the present model is completely negligible. This is very welcome in view of the good agreement of $Br(B\to X_s\gamma)$ with the data. 

Contributions to the four-fermion operators mediating transitions like $b\to s\mu^+\mu^-$ or $s\to d\nu\bar\nu$ can generally be split into tree level and one loop box and penguin diagrams. In the mDMFV model new tree level diagrams are forbidden by the residual $\mathbb{Z}_3$ symmetry (see appendix \ref{app:Z3}), while box diagrams are not generated since the new particles $\phi$ and $\chi_i$ do not couple to leptons. We are hence left with potential contributions to the $Z$ and photon penguins. An explicit calculation shows that the $Z$ penguin contribution vanishes. This can be explained by the chiral structure of our model with the new particles coupling only to right-handed quarks, and is also confirmed by adapting the SUSY results of  \cite{Cho:1996we} to our scenario.
The photon penguin contribution is non-zero, however numerically small, as known from supersymmetric models \cite{Cho:1996we,Altmannshofer:2013foa}. 

In summary we are left with completely SM-like rare decays like $B_{s,d}\to\mu^+\mu^-$, $B\to K^*\mu^+\mu^-$ and $B\to X_s\gamma$. Consequently the mDMFV model does not ameliorate the tension in the $B\to K^*\mu^+\mu^-$ data.

A bit more care is however required in the case of semileptonic decays
with neutrinos in the final state, such as $K\to\pi\nu\bar\nu$ or
$B\to K^{(*)}\nu\bar\nu$. Since the neutrinos escape detection, the
experimental signatures are $K\to\pi+\EmissT$ and $B\to
K^{(*)}+\EmissT$ respectively. Consequently also the
decays\footnote{We denote by $\chi_\text{DM}$ the lightest flavor
which is stable and provides the DM.} $K\to\pi\chi_\text{DM}\bar\chi_\text{DM}$ and $B\to K^{(*)}\chi_\text{DM}\bar\chi_\text{DM}$, mediated by a tree level $\phi$ exchange, will contribute to the measured branching ratio if the decay is kinematically allowed. See \cite{Kamenik:2011vy} for a detailed discussion. In order to avoid these potentially stringent constraints, in the remainder of our analysis we will assume $m_\text{DM} > 10\gev$ and therefore well outside the kinematically allowed region for these decays.

\subsection{Electroweak precision tests and electric dipole moments}

Besides the flavor violating $K$ and $B$ decays discussed above,
the flavor conserving electroweak precision constraints and the bounds
on electric dipole moments also put strong constraints on many NP models.
In this section we consider these observables within the mDMFV model.

We start by considering electroweak precision observables. Due to the
residual $\mathbb{Z}_3$ symmetry corrections from the mDMFV model can
arise only at the loop level and are therefore suppressed by a loop
factor $1/(16\pi^2)$. Furthermore the mDMFV model introduces no new
$SU(2)_L$ doublets, and only $\phi$ carries hypercharge. Consequently
the contributions to electroweak precision observables receive an
additional suppression by $\sim g_Y^2/(9 m_\phi^2)$. Together with the
loop factor and the scale $m_\phi$ above the electroweak scale we
conclude that all new contributions to electroweak precision
observables are safely small.

Similarly we also find no significant new contribution to electric
dipole moments. The reasons are as follows. Due to the chiral
structure of the mDMFV model with new particles coupling only to
right-handed down-type quarks no EDM is generated at the one loop
level. At the two loop level a Barr-Zee type diagram
\cite{Barr:1990vd} with $\phi$ running in the loop exists -- however
its CP-violating phase is zero because the coupling
$\lambda_{H\phi}$ is real.

\section{Flavor pre-analysis of possible structures for the DM-quark coupling}\label{sec:DF2numerics}

We are now prepared to study the allowed regions of parameter space from flavor observables as well as correlations between different parameters of the coupling matrix $\lambda$. 

\subsection{Strategy of the numerical analysis}

In order to determine the constraints from $\Delta F = 2$ observables
on the mDMFV model, we use the latest New Physics Fit results of the
model-independent NP fit presented by the UTfit collaboration
\cite{Bona:2005eu}. To this end we define
\be
M_{12}^{B_q} = C_{B_q} e^{2i \varphi_{B_q}} M_{12}^{B_q,\text{SM}}\qquad (q=d,s)\,,
\ee
where $M_{12}^{B_q}$ is the full mixing amplitude containing both SM and mDMFV contributions. Furthermore
\be
\text{Re} M_{12}^{K} = C_{\Delta M_K}  \text{Re} M_{12}^{K,\text{SM}}\,, \qquad \text{Im} M_{12}^{K} = C_{\varepsilon_K}  \text{Im} M_{12}^{K,\text{SM}}\,.
\ee
These six parameters are constrained by a global fit of the NP amplitude to the available tree level and $\Delta F = 2$ data \cite{Bona:2005eu,Bona:2007vi}. In order to be conservative we impose the resulting constraints at the $2\sigma$ level, see table \ref{tab:inputs} for a summary. In the case of $\Delta M_K$ we allow for a $\pm40\%$ uncertainty in order to capture the poorly known long distance effects. For consistency we set the CKM parameters to their central values obtained in the UTfit fit. All other input parameters are set to their central values listed in table 3 of \cite{Buras:2013dea}.
\begin{table}[h!]
\centering{
\begin{tabular}{|l|l|}
\hline
$|V_{us}|=0.22527$ & $C_{\Delta M_K}=1.10\pm0.44$\\
$|V_{ub}|=3.76\cdot 10^{-3}$ & $C_{\varepsilon_K}=1.05 \pm 0.32$\\
$|V_{cb}|=4.061\cdot 10^{-2}$ & $C_{B_d}=1.07 \pm 0.34$\\
$\delta= 67.8^\circ$ & $\varphi_{B_d}=-(2.0 \pm 6.4)^\circ$\\
& $C_{B_s}=1.066 \pm 0.166$\\
& $\varphi_{B_s}=(0.6 \pm 4.0)^\circ$ \\\hline
\end{tabular}}
\caption{\label{tab:inputs}Summary of CKM parameters and $\Delta F=2$ constraints used in our numerical analysis, see \cite{Bona:2005eu} for details.}
\end{table}

Altogether, we have the following new parameters relevant for flavor
violating decays:
\be
m_\phi\,,\quad m_\chi\,,\quad \lambda_0\,,\quad \lambda_1\,,\quad
\lambda_2\,,\quad \theta_{12}^\lambda\,,\quad \theta_{13}^\lambda
\,,\quad \theta_{23}^\lambda\,,\quad \delta_{12}^\lambda\,,\quad
\delta_{13}^\lambda \,,\quad \delta_{23}^\lambda \,.
\ee

Our goal is to obtain a clear picture of patterns in the coupling matrix $\lambda$ that are implied by the available $\Delta F = 2$ data. To this end we fix the flavor conserving parameters $m_\phi$, $m_\chi$ and $\lambda_0$ to the values
\be
m_\phi=850\gev\,,\qquad m_\chi= 200\gev\,,\qquad
\lambda_0=1\,.
\ee
The impact of varying these parameters can be estimated from the functional dependence of $M_{12}^{i,\text{new}}$ ($i=K,B_d,B_s$), which is roughly given by
\be
M_{12}^{i,\text{new}} \propto \frac{\lambda_0^2}{m_\phi^2} F(x)\,,\qquad x= m_\chi^2/m_\phi^2\,.
\ee
Note that $F(x)$ is a monotonically decreasing function varying from 1 to 1/3 over the range $0<x<1$.

\begin{figure}
\centering
\includegraphics[width=.6\textwidth]{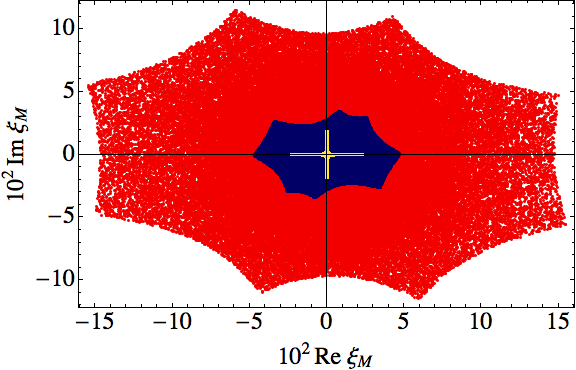}
\caption{\label{fig:xi}Allowed ranges for the flavor violating parameters $\xi_M=\xi_K$ (yellow), $\xi_M=\xi_{B_d}$ (blue), $\xi_M=\xi_{B_s}$ (red).}
\end{figure}

The $\Delta F =2$ constraints then translate directly into constraints on the values of $\xi_K$, $\xi_{B_d}$ and $\xi_{B_s}$, as shown in figure \ref{fig:xi}.
We observe that the strongest constraints come from $K^0 -\bar K^0$ mixing, and in particular the CP-violating parameter $\varepsilon_K$, which forces the phase of $\xi_K$ to be very close to $0,\pi/2,\pi,3\pi/2$ unless $|\xi_K|\lsim 10^{-3}$. The $B$ physics constraints are less stringent and in particular do not yield a specific pattern for the phases of $\xi_{B_q}$. The weakest constraints are found in the $B_s$ system.

This pattern of allowed deviations from the SM is not specific to the mDMFV model, but can be found in all models with a generic NP flavor structure that do not induce the chirally enhanced left-right operators, like the Littlest Higgs model with T-parity analyzed in detail in \cite{Hubisz:2005bd,Blanke:2006sb,Blanke:2009am}. It is a direct consequence of the CKM hierarchies that determine the size of effects within the SM, as well as the theoretical uncertainties involved.
Note that e.\,g.\ in Randall-Sundrum (RS) models with bulk fermions \cite{Csaki:2008zd,Blanke:2008zb,Bauer:2009cf} and left-right models \cite{Zhang:2007da,Blanke:2011ry}, the strong enhancement of the left-right operators in the kaon system makes the $K^0-\bar K^0$ constraints even more severe.

\subsection[``Flavor-safe'' scenarios for the structure of $\lambda$]{\boldmath ``Flavor-safe'' scenarios for the structure of $\lambda$}

We now analyze the structure of the coupling matrix $\lambda$ that is implied by the $\Delta F =2 $ constraints. To this end we show in figure \ref{fig:scenarios}
 the allowed points in the $(\lambda_1,\lambda_2,s_{12}^\lambda)$, $(\lambda_1,\lambda_2,s_{13}^\lambda)$ and $(\lambda_1,\lambda_2,s_{23}^\lambda)$ spaces, respectively. 
 \begin{figure}[htp]
\centering{
\includegraphics[width=.6\textwidth]{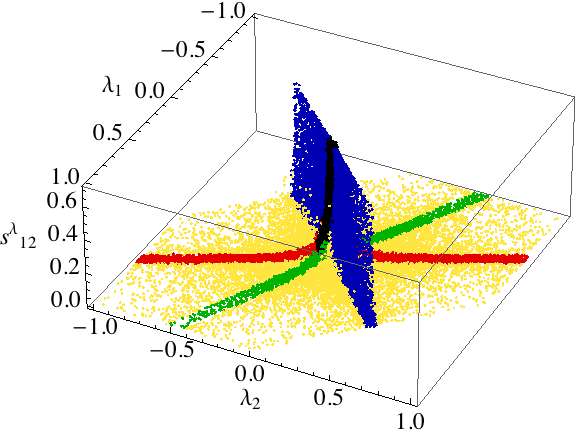}

\includegraphics[width=.6\textwidth]{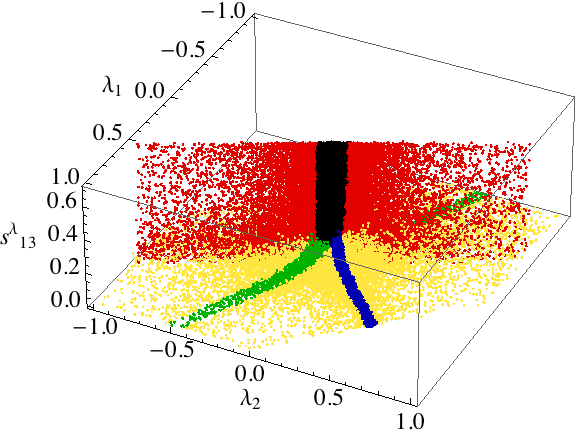}

\includegraphics[width=.6\textwidth]{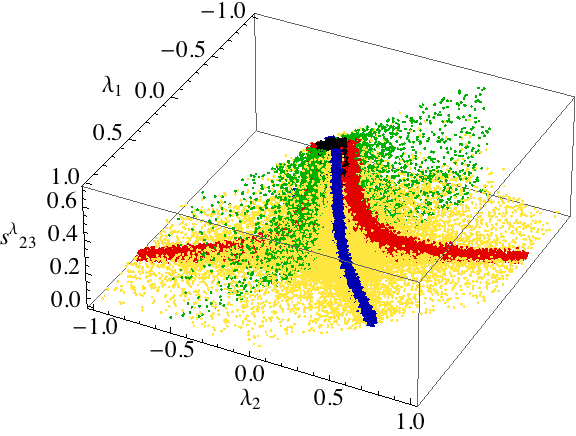}}
\caption{\label{fig:scenarios}Scenarios for the structure of $\lambda$: Universality (black), 12-degeneracy (blue), 13-degeneracy (red), 23-degeneracy (green), 
small mixing (yellow).}
\end{figure}
\afterpage{\clearpage}

We observe that the allowed points fall into five distinct scenarios for the structure of $\lambda$, which we discuss in some detail in the following. In order to analytically understand the scenarios, we recall the parametrization of $\lambda$ in terms of three two-flavor rotation matrices $U_{ij}$ and a diagonal matrix $D_\lambda$:
\be
\lambda = U_{23} U_{13} U_{12} D_\lambda\,.
\ee

\begin{enumerate}
\item {\bf universality scenario} (black): $\lambda_1 \simeq \lambda_2  \simeq 0 $

In this case $\lambda \simeq U_\lambda \cdot \lambda_0$ so that $\lambda \lambda^\dagger \simeq \lambda_0^2\cdot\mathbbm{1}$. Since flavour violating effects are governed by the off-diagonal elements of $\lambda \lambda^\dagger$, the $\Delta F=2$ constraints are trivially fulfilled for arbitrary $U_\lambda$ and there are no FCNC effects beyond the SM.

\item {\bf 12-degeneracy} (blue): $\lambda_1 \simeq \lambda_2$

If the first two generations of DM fermions are quasi-degenerate, then -- as seen from the blue points in figure \ref{fig:scenarios} -- the mixing angle $s_{12}^\lambda$ can be generic while $s^\lambda_{13,23}$ have to be small. This can be understood by taking the limit $\lambda_1=\lambda_2$, in which the mixing matrix $U_{12}$ becomes non-physical, and we are left with 
\begin{equation}
\lambda= U_{23} U_{13} D_\lambda\,.
\end{equation}
It is easy to see that in order to fully suppress flavor violating effects we need $U_{13,23}\simeq \mathbbm{1}$ and therefore $s^\lambda_{13,23}\simeq 0$.

\item {\bf 13-degeneracy} (red): $\lambda_2 \simeq -2\lambda_1$

In the case $\lambda_2 \simeq -2\lambda_1$, shown by the red points in figure \ref{fig:scenarios}, the first and third DM flavor are quasi-degenerate, and consequently $s_{13}^\lambda$ is unconstrained. In order to suppress the remaining flavor violating effects both $s^\lambda_{12}$ and $s^\lambda_{23}$ have to be small.

\item {\bf 23-degeneracy} (green): $\lambda_2 \simeq -1/2\lambda_1$

Finally if $\lambda_2 \simeq -1/2\lambda_1$, the second and third DM flavor are quasi degenerate. Consequently the mixing angle $s^\lambda_{23}$ is arbitrary, while $s^\lambda_{12}$ and $s^{\lambda}_{13}$ have to be small.
This scenario is shown by the green points in figure \ref{fig:scenarios}.

\item
{\bf small mixing scenario} (yellow): arbitrary $D_\lambda$

Finally if $D_\lambda$ does not exhibit any degeneracies, then FCNC effects have to be suppressed by the smallness of all three mixing angles $s_{12}^\lambda \simeq s_{13}^\lambda \simeq s_{23}^\lambda \simeq 0$. This scenario, shown by the yellow points in figure \ref{fig:scenarios}, corresponds to a diagonal but non-degenerate coupling matrix $\lambda$. 
\end{enumerate}

In order to quantify the allowed size of deviations from the
degeneracy scenarios discussed above, we show in figure \ref{fig:deg}
the mixing angles $s^{\lambda}_{ij}$ as a function of the deviation
from the corresponding degeneracy line. We observe that the constraint
on $12$  and $13$ mixings are comparable with the former being
somewhat stronger, while the constraint on the $23$ mixing angle is
significantly weaker. This is a direct consequence of the allowed
sizes of NP effects in the various meson systems, see figure
\ref{fig:xi}.

\begin{figure}[h!]
\centering
\includegraphics[width=.6\textwidth]{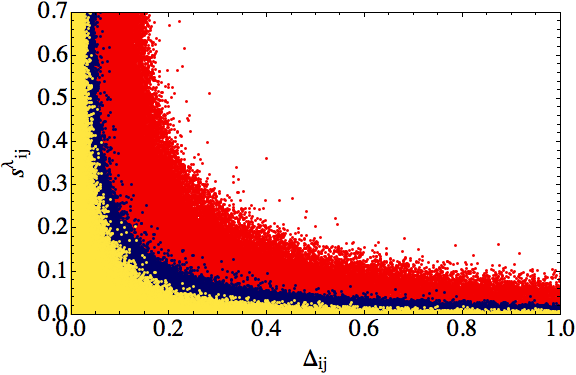}
\caption{\label{fig:deg}Allowed ranges for the mixing angles $s^\lambda_{ij}$ as a function of the deviation from the $ij$-degeneracy line $\Delta_{ij}=|D_{\lambda,ii}-D_{\lambda,jj}|$. $ij=12$ in yellow, $ij=13$ in blue, $ij=23$ in red.}
\end{figure}

\subsection{A note on flavor safety of the UV completion}

FCNC processes are known to be sensitive to NP at very high scales. It
is therefore questionable whether a study of the simplified mDMFV
model is sufficient to capture all relevant effects.

Following the DMFV principle we can write any contribution from the UV
completion in terms of higher-dimensional operators that are
suppressed by powers of the UV scale $\Lambda_\text{UV}$ and made formally
invariant under the flavor group \eqref{eq:flavor-group} by insertion
of the appropriate combination of spurion fields $Y_{u,d}$ and
$\lambda$. The leading contribution to the $\Delta F = 2$ effective
Hamiltonian is then
\be
\Heff^{\Delta F=2,\text{UV}} \sim
\frac{c^\text{UV}_{\Delta F=2}}{\Lambda_\text{UV}^2} \lambda \lambda^\dagger  (\bar s \gamma^\mu P_R d) (\bar s \gamma_\mu P_R d)\,,
\ee
where $c^\text{UV}_{\Delta F=2}$ is an $\ord(1)$ coefficient that is common to all three meson systems. Comparing this to the new contribution generated first at the one loop level in the simplified model (see \eqref{eq:Heff1}), that can schematically be written as
\be
\Heff^{\Delta F=2, \text{simpl.}} \sim
\frac{c^\text{simpl.}_{\Delta F=2}}{16\pi^2 m_\phi^2} \lambda \lambda^\dagger  (\bar s \gamma^\mu P_R d) (\bar s \gamma_\mu P_R d)\,,
\ee
we observe that both contributions carry the same flavor structure. Furthermore the UV contribution is suppressed with respect to the simplified model one if
\be
\Lambda_\text{UV}\gsim 4\pi m_\phi\,.
\ee

Therefore, NP close to the mass of $\phi$ does not change the
flavor phenomenology as long as it respects the DMFV hypothesis.
Generic flavor violation needs to be suppressed by a much higher scale
$\sim\mathcal{O}({100-1000})$ TeV~\cite{Isidori:2010kg}.

\subsection[Recovering the MFV limit in the structures for $\lambda$]{\boldmath Recovering the MFV limit in the structures for $\lambda$}

Earlier studies of flavored DM have been restricted to the MFV
framework in order to be safe from undesired effects in flavor
observables.  The flavor-safe scenarios identified above
are more general than the MFV ansatz. It is also worthwhile to study
how MFV can be recovered in the DMFV framework. 

Let us first consider the case where $U(3)_\chi$ is identified with $U(3)_d$. The MFV hypothesis then requires that $\lambda$ takes the schematic form
\be\label{eq:lambdaMFV}
\lambda
\propto \mathbbm{1} +\alpha Y_d^\dagger Y_d + \dots\,,
\ee
where $\alpha$ is an arbitrary coefficient.
The matrix $\lambda$ is diagonal in the down quark mass basis. In particular $Y_d$ is
proportional to the down-type quark masses. Considering that $Y_d$ can
be approximated by $Y_d\sim \diag(0,0,y_b)$, MFV must be close to the
12-degeneracy. Additionally MFV requires 
\be 
m_\chi \propto \mathbbm{1}+ \beta Y_d^\dagger Y_d + \dots\,,
\ee
where $\beta$ is an arbitrary coefficient.
The same expansion for $m_\chi$ is obtained when inserting \eqref{eq:lambdaMFV} into the DMFV expansion \eqref{eq:DMFVmass}, so that MFV in this case is consistent with the DMFV hypothesis. As $\lambda$ and $m_\chi$ are diagonal in the same basis, all three flavor mixing angles are zero. Thus the MFV limit can be recovered as a very specific subset of parameter space, (determined by the specific choice of $Y_d$) close to the 12-degeneracy line with all mixing angles zero.

If instead  $U(3)_\chi$ is identified with $U(3)_u$, then 
\be
\lambda
\propto Y_d^\dagger Y_u + \dots\,,
\ee
while
\be 
m_\chi \propto \mathbbm{1}+ \alpha Y_u^\dagger Y_u + \dots\, ,
\ee
where as before $\alpha$ is an arbitrary coefficient.
 We can see immediately that the mass splittings are not
directly
correlated with the $\lambda$ matrix, as was the case for
DMFV \eqref{eq:DMFVmass} with the separate U(3)$_\chi$ symmetry.

Finally identifying $U(3)_\chi$ with $U(3)_q$, we have
\be
\lambda
\propto Y_d + \dots
\ee
and
\be 
m_\chi \propto \mathbbm{1}+ \alpha Y_u Y_u^\dagger + \beta Y_d Y_d^\dagger + \dots\,,
\ee
where $\alpha$ and $\beta$ are arbitrary coefficients.
 Again we observe that the pattern of splittings in $m_\chi$ 
are not directly correlated with the $\lambda$ matrix.

In summary we find that if $\chi$ is assumed to transform under
$U(3)_d$ then the MFV limit can be recovered as a small subset of the
scenarios for $\lambda$, with an approximate 12-degeneracy and all
mixing angles identically zero.  On the other hand, the
MFV limits for
$\chi$ transforming under $U(3)_u$ or $U(3)_q$ have mass splitting
patterns which are not correlated with the coupling matrix $\lambda$,
and therefore to capture these cases one needs to consider additional
contributions to the mass splitting in \eqref{eq:DMFVmass}.

\section{From the flavor pre-analysis to dark matter scenarios}\label{sec:flavor-consequences}

Our flavor pre-analysis shows that a generic coupling matrix $\lambda$
leads to unacceptably large corrections to $\Delta F = 2$ observables.
We have identified a number of non-trivial scenarios for the structure
of $\lambda$ for which flavor violating effects are efficiently
suppressed:
\begin{enumerate}
\item
Universality scenario: all elements of the diagonal matrix $D_\lambda$ equal and arbitrary flavor mixing angles.
\item
$ij$-degeneracy scenarios ($ij=12,13,23$): $D_{\lambda,ii}=D_{\lambda,jj}$, arbitrary $s^\lambda_{ij}$ and the other mixing angles small.
\item
Small mixing scenario: small mixing angles and arbitrary $D_\lambda$.
\end{enumerate}

While these scenarios have been identified in a scan with fixed flavor
conserving parameters $m_\phi$, $m_\chi$ and $\lambda_0$, we stress
that these structures for $\lambda$ also remain valid for different
choices of parameters. Furthermore, even though our analysis has been
performed within the simplified framework of the mDMFV model, the
identified scenarios for $\lambda$ remain flavor-safe in
non-minimal versions of DMFV also. Thus they provide a useful framework for
future study of the phenomenology of DMFV models -- employing any of
these scenarios for the structure of $\lambda$ efficiently evades all
FCNC constraints, without the need for an involved study of the
latter.

\begin{table}[h!]
  \begin{center}
    \begin{tabular}{|l|l|l|}
      \hline
  scenario &  specification  & lightest DM particle \\ \hline\hline
    universal scenario ($ m_{\chi_d} \simeq m_{\chi_s} \simeq m_{\chi_b}$) & -  &  all hierarchies possible  \\  \hline
    12 degeneracy ($m_{\chi_d} \simeq m_{\chi_s}$)  & $\eta \lambda_1 > 0$  &   $ {\chi_b}$  \\
    & $\eta \lambda_1 < 0$  &   ${\chi_d}$ or ${\chi_s} $  \\  \hline
13 degeneracy ($m_{\chi_d} \simeq m_{\chi_b}$)  & $\eta \lambda_1 > 0$   &   $ {\chi_s}$  \\
    & $\eta \lambda_1 < 0$   &   ${\chi_d}$ or ${\chi_b} $  \\ \hline
23 degeneracy  ($m_{\chi_s} \simeq m_{\chi_b}$)  & $\eta \lambda_1 > 0$  &   ${\chi_s}$ or ${\chi_b}$ \\
    & $\eta \lambda_1 < 0$   &   ${\chi_d}$  \\
     \hline
small mixing scenario  &  -  &   all hierarchies possible \\  \hline
    \end{tabular}
  \end{center}
  \caption{Overview of flavor-safe scenarios and their implications
  for the mass hierarchy in the DM sector. \label{tab:DMscenarios}}
\end{table} 

In table \ref{tab:DMscenarios} we summarize the flavor-safe scenarios
for $\lambda$ and their implications for the mass pattern in the DM
sector. It is clear that flavor constraints do not impose a
specific mass hierarchy on the dark sector, i.\,e.\ from the point of
FCNC constraints any dark flavor can be the lightest. Note that an
exact degeneracy of two flavors is unnatural, since in the case of
universal $\lambda$ it is violated by the presence of $Y_{d,u}$ at
higher orders in the DMFV expansion. We therefore assume that the
observed DM is composed of a single $\chi$ flavor, while the
decay of the heavier states is fast enough to have happened in the
early universe. We refer the reader to appendix \ref{sec:decayheavy} for an
estimate of the life-time of the heavier states. 

However not all DM flavors are equally motivated from the
point of DM and collider phenomenology. DM that
couples dominantly to first generation quarks, like $d$-flavored DM, 
is strongly constrained by the direct detection experiments.
If the DM relic density is assumed to arise from thermal
freeze-out in the early universe, the relic abundance condition is in
severe tension with the experimental constraints. We will therefore
not consider the case of $d$-flavored DM further.

As far as direct detection constraints are concerned, $s$- and
$b$-flavored DM are on equal footing.  Interestingly the same
holds, at least qualitatively, also for the flavor phenomenology -- as we have seen in figure
\ref{fig:deg} the amount of flavor violation allowed by $\Delta F=2$
constraints is almost symmetric under the exchange of the second and
third generation, $2\leftrightarrow 3$.

The case is however different for collider phenomenology. While pair
production of the mediator and its subsequent decay will dominantly
produce light jets and missing energy in the $s$-flavored case, in the
case of $b$-flavored DM the large coupling to the $b$ quark
will give rise to $b$-jet signatures in a significant fraction of the
events. Since events with $b$-jets are much more easily distinguished
from the QCD background, the collider phenomenology of $b$-flavored
DM is at the same time more constraining (in particular
concerning the bound on the mediator mass) and also more promising, as
quite distinctive 
signatures arise. 

A further motivation for $b$-flavored DM comes from indirect
detection. Recently it has been shown that a 35 GeV $\chi_b$ provides
a good fit to the excess $\gamma$-rays observed at
the galactic center \cite{Agrawal:2014una}.

Therefore, in the rest of our analysis we restrict ourselves to the
case of $b$-flavored DM, i.\,e.\ $m_{\chi_b}<m_{\chi_{d,s}}$.
We also assume that the DM relic abundance is set by the
thermal freeze-out condition, so that $D_{\lambda,33}$ has to be
large. Due to the strong constraints on the first generation coupling
from direct detection and collider data, we deduce that
$D_{\lambda,11}<D_{\lambda,33}$. Consequently in order to ensure the
correct mass hierarchy, we have $\eta < 0$. 

We are then left with the following scenarios for DM freeze-out:
\begin{enumerate}
  \item single flavor freeze-out: The $s$- and $d$-flavored states are split from the $b$-flavored DM by at least 10\%.
  \item two flavor freeze-out: 
    \begin{enumerate}
      \item 13-degeneracy -- $\chi_b$ and $\chi_d$ are quasi-degenerate, while $\chi_s$ is split
      \item 23-degeneracy -- $\chi_b$ and $\chi_s$ are quasi-degenerate, while $\chi_d$ is split
    \end{enumerate}
  \item three flavor freeze-out: 
    All three states are quasi-degenerate. Such a scenario can either be achieved by a quasi universal coupling matrix $D_\lambda$, or if the DMFV expansion parameter $\eta$ is loop-suppressed, $|\eta|\sim 10^{-2}$.
\end{enumerate}
In our numerical analysis we will study all of these scenarios in turn.

\section{\boldmath Phenomenology of $b$-flavored dark matter}
\label{sec:DMpheno}

In this section, we study the constraints arising from requiring the
DM to be a thermal relic and from direct
detection experiments. We note that the relic abundance constraints
may be potentially relaxed in the presence of other particles in the
dark sector.

The presence of multiple flavors can affect the DM freeze-out
significantly. This occurs when the mass splitting between different
flavors of DM is much smaller than the freeze-out
temperature, ($T_f \sim m_\chi / 20$). For mass splittings much bigger
than this scale, the freeze-out follows the standard WIMP paradigm.

We show that in the range of parameter space we consider, the heavier
DM flavors decay before big bang nucleosynthesis (BBN) (see appendix
\ref{sec:decayheavy}).
The direct detection constraints then depend sensitively on the
couplings of the lightest flavor of DM. In particular, when
the lightest dark flavor couples appreciably to the first generation quarks,
it gives rise to a very large direct detection signal. If this
contribution is suppressed, the dominant contribution then arises at
1-loop level, which is seen to be within the reach of present and
future direct detection experiments.

\subsection{Relic abundance}

\begin{figure}[h!]
  \centering
  \includegraphics[width=0.3\textwidth]{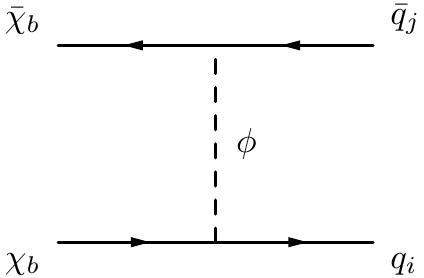}
  \caption{Feynman diagram for dark matter annihilation in the early universe.}
  \label{fig:relic}
\end{figure}

We will consider two different qualitative regimes. When their masses
are nearly degenerate, then all DM flavors are present during
freeze-out and can be treated together. Otherwise only the lightest
flavor of DM remains in the thermal
bath. 

We start with a single flavor freeze-out.
The dominant annihilation during freeze-out occurs in the lowest
partial wave. In
this limit
\begin{align}
  \langle\sigma v\rangle_{bb}
  &=
  \sum_{i,j}
  \frac{3\lambda_{ib} \lambda^*_{ib}
  \lambda_{jb}\lambda^*_{jb} m_{\chi_b}^2}{32\pi (m_{\chi_b}^2+m_\phi^2)^2}
  =
  \frac{3(D_{\lambda}^{\dagger}  U^\dagger  U D_\lambda)^2_{33}
  m_{\chi_b}^2}{32\pi (m_{\chi_b}^2+m_\phi^2)^2}
  =
  \frac{3D_{33}^4
  m_{\chi_b}^2}{32\pi (m_{\chi_b}^2+m_\phi^2)^2},
  \label{annihilationcrosssection}
\end{align}
where we have ignored the masses of the final state quarks.

The relic abundance is determined by solving the Boltzmann equation for
the DM number density $n$ at late times.
For a Dirac fermion, it is useful to convert the annihilation cross
section into an effective cross
section~\cite{Griest:1990kh,Servant:2002aq}.
\begin{align}
\langle\sigma v\rangle_\text{eff} = \frac12 \langle
\sigma v\rangle \ .
\end{align}
which is approximately required to be~\cite{Steigman:2012nb}
\begin{align}
  \langle \sigma v \rangle_\text{eff}
  &=
  2.2\times10^{-26 }\mathrm{cm}^3/\mathrm{s}
  \label{eq:effsigma}
\end{align}
in order to produce the correct relic abundance of DM.

Next we consider the case, where the mass
splitting between the DM flavors is much smaller than the
temperature at freeze-out. Consequently, we have to take into account
the co-annihilation between different flavors.  We assume that flavor
changing (but DM number preserving) interactions $\chi_i q
\to \chi_j q$ are fast during the epoch of DM freeze-out. The
rate for these processes is enhanced over the DM
annihilations---which are approximately in thermal equilibrium---by a
large Boltzmann factor ($\mathcal{O}(10^9)$). Thus, this approximation
is valid as long as any individual cross sections are not suppressed
enough to overwhelm this factor.

Then the Boltzmann equation for freeze-out
has a very similar form to the single DM
case, and can be solved in exactly the same way.  The
relic abundance is in fact relatively insensitive to the change in the
number of DM species, changing by only about 5\% when other
parameters are kept fixed.  In the limit of small splitting,
the effective cross section 
is well approximated by \cite{Griest:1990kh},
\begin{align}
  \langle \sigma v \rangle_\text{eff}
  &=
  \frac{1}{18}
  \sum_{i,j=d,s,b}
  \langle \sigma v \rangle_{ij}
  .
\end{align}
The co-annihilation cross section can be derived by modifying equation
(\ref{annihilationcrosssection}), e.g.
\begin{align}
  \langle\sigma v\rangle_{bs}
  &=
  \sum_{i,j}
  \frac{3 \lambda_{is} \lambda^*_{is}
  \lambda_{jb}\lambda^*_{jb} m_{\chi b}^2}{32\pi (m_{\chi b}^2+m_\phi^2)^2}
  =
  \frac{3 D_{22}^2 D_{33}^2
  m_{\chi b}^2}{32\pi (m_{\chi b}^2+m_\phi^2)^2}\,.
  \label{annihilationcrosssectiontwofl}
\end{align}
Note that the splitting between DM masses is in this case
negligible ($m_{\chi d} \simeq m_{\chi s} \simeq m_{\chi b}$).

If only two states are nearly degenerate, then a two flavor freeze-out occurs. The corresponding formulae can be straightforwardly obtained from the above results.

\subsection{Direct detection}

For $b$-flavored DM, the
direct detection scattering arises either through mixing, or at one
loop.

We focus on the spin-independent contribution to the WIMP-nucleus
scattering. The reported experimental bounds are translated to the 
WIMP-nucleon cross section, which can be written as
{\begin{align}
  \sigma^{\rm SI}_n
  &= 
  \frac{\mu^2_n}{\pi A^2}
  \left| Z f_p + (A-Z)f_n\right|^2\,,
\end{align}}
where $\mu_n$ is the reduced mass of the WIMP-nucleon system,
$A$ and $Z$ are the mass and atomic numbers of the nucleus
respectively, and $f_n$ and $f_p$ parametrize DM coupling to
neutrons and protons. The relevant processes are shown in figure \ref{fig:dd}.
\begin{figure}[h!]
  \includegraphics[width=0.3\textwidth]{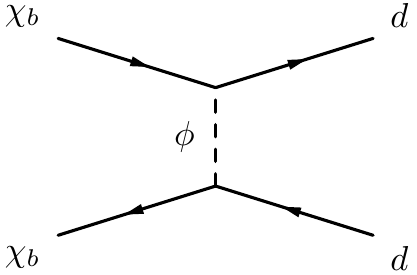}\hfill
  \includegraphics[width=0.3\textwidth]{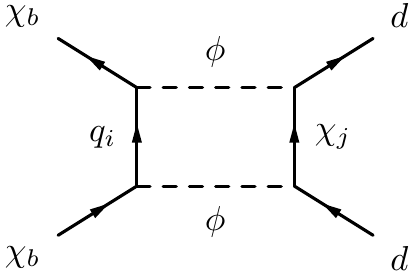}\hfill
  \includegraphics[width=0.3\textwidth]{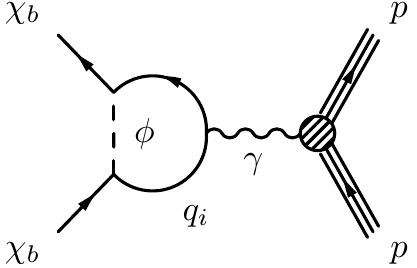}
  \caption{Diagrams contributing to WIMP-nucleon scattering in the mDMFV model.}
  \label{fig:dd}
\end{figure}

For direct detection, we can safely work in the effective theory with
the $\phi$ integrated out. The Lorentz structure of the 
four-fermion operator generated after
performing the Fierz transformation is given by
\begin{align}
\bar{\chi}_b \gamma^\mu (1-\gamma^5)\chi_b\,
\bar{d} \gamma^\mu (1+\gamma^5)d\,.
\label{eq:effoper}
\end{align}
There are three contributions to $f_{n,p}$:
\begin{align}
  f_{n,p} &= 
  f^{\rm tree}_{n,p} +
  f^{\rm box}_{n,p} +
  f^{\rm photon}_{n,p} \,.
\end{align}
These contributions are individually given as follows:
\begin{enumerate}
  \item $s$-channel $\phi$ at tree-level:\\
    In presence of significant mixing in the $\lambda$ matrix, 
    this is the dominant contribution to direct detection:
    \begin{align}
      2f^{\rm tree}_p
      =f^{\rm tree}_n = \frac{ |\lambda_{db}|^2}{4 m_\phi^2}\,.
    \end{align}
    The spin-independent part in equation (\ref{eq:effoper}) arises
    from the matrix element of the quark vector current bilinear 
    in the nucleons. Thus, 
    only the valence $d$-quark contributes.
  \item One-loop photon exchange:\\
    The interaction of DM with nucleons via photon exchange
    is conveniently parametrized as the electromagnetic form factors
    of the DM coupling with those of the nucleus.
    In particular, the scattering cross sections
    arise from charge-charge, dipole-charge and dipole-dipole
    interactions~\cite{Agrawal:2011ze}. In the region of interest, the
    charge-charge
    interactions dominate, leading to
    \begin{align}
      f^\text{photon}_p
      &=
      \sum_i
      \frac{\left| \lambda_{ib} \right|^2 e^2} 
      {96 \pi^2 m_\phi^2}
      \log \left[\frac{m_{q_i}^2 }{ m_\phi^2}\right] 
    \end{align}
    in the leading-log approximation.
  \item Box diagram with $\phi$ exchange in the t-channel:\\
    {This new contribution depends upon} the coupling of DM with the first
    generations quarks \cite{Kumar:2013hfa} and is given by
    \begin{align}
      4 f_p^{\rm box} &= 
      2 f_n^{\rm box} = 
      \sum_{i,j} \frac{\left| \lambda_{dj} \right|^2 \left| \lambda_{ib}\right|^2 } {16\pi^2 m_\phi^2}
      F \left( \frac{m_{q_i}^2}{m_\phi^2},
      \frac{m_{\chi_j}^2}{m_\phi^2}\right)\,,
    \end{align}
    with the loop function $F$ given in appendix \ref{app:functions}.
    
\end{enumerate} 
The tree-level contribution, {being flavor violating}, constrains the mixing $s^\lambda_{13}$ to be
small. The LUX experiment~\cite{Akerib:2013tjd} is sensitive to even the loop level
scattering cross sections for WIMP DM. {These contributions are present even in the absence of flavor
   violation.} The box and the photon
loop diagrams are seen to destructively interfere.

\section{Combined numerical analysis of flavor and dark matter constraints}\label{sec:numerics}

Having all relevant formulae for the DM phenomenology in
hand, we are now ready to perform a combined numerical analysis of
both DM and flavor constraints. We restrict ourselves to the
phenomenologically most interesting case of $b$-flavored DM,
and study in turn the scenarios identified in section
\ref{sec:flavor-consequences}.

The DM mass $m_{\chi_b}$ is allowed to vary in the
phenomenologically interesting region $10\gev<m_{\chi_b}<250\gev$.  We
also assume $\eta<0$ in order to suppress the $\chi$ coupling to the
first generation, in order to cope with the strong direct detection
and collider constraints. Convergence of the DMFV expansion is ensured
by requiring $|\eta D_{\lambda,33}^2|<0.3$. Finally, since corrections
to $m_\chi$ are unavoidably generated at the one loop level, we take
$|\eta|>10^{-2}$. In summary, \be
-\frac{0.3}{D_{\lambda,33}^2}<\eta<-0.01\,.  \ee We fix
$m_\phi=850\gev$ in agreement with the collider constraints, see
section \ref{sec:collider}.  The parameters of the coupling matrix
$\lambda$ are scattered, imposing the flavor and DM constraints from
section \ref{sec:DF2constraints} and \ref{sec:DMpheno}.

\subsection{Single flavor freeze-out}

If the masses of the heavier flavors $\chi_{d,s}$ are sufficiently
split from the DM mass $m_{\chi_b}$ (by $\gsim 10\%$), then at the freeze-out temperature only the lightest state $\chi_b$ is left while the heavier ones have decayed. Therefore the single flavor freeze-out condition for the relic abundance, eq.\ \eqref{eq:effsigma}, applies and fixes $D_{\lambda,33}$ as a function of $m_{\chi_b}$.
The couplings $D_{\lambda,11}$ and $D_{\lambda,22}$ on the other hand are free to vary. They  both need to be split from $D_{\lambda,33}$ in order to achieve the mass splitting.

\begin{figure}[h!]
\includegraphics[width=.49\textwidth]{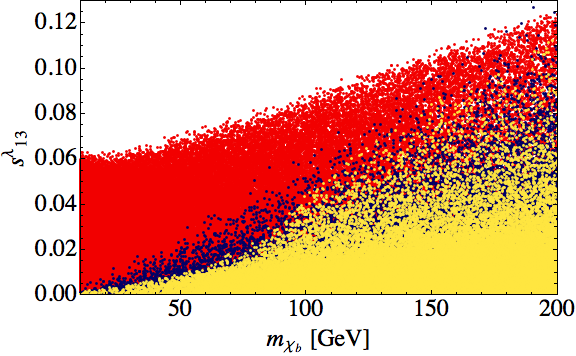}\hfill
\includegraphics[width=.49\textwidth]{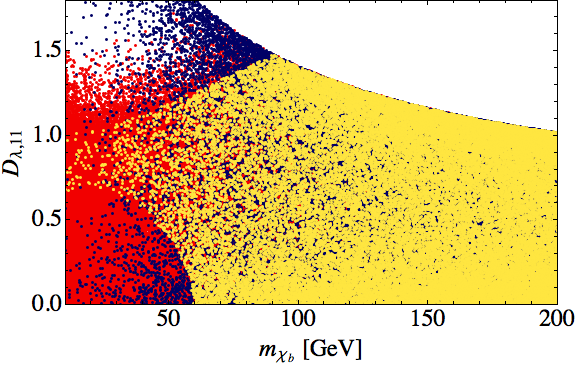}
\caption{\label{fig:1f-s13-D11}Flavor mixing angle $s^\lambda_{13}$
and first generation coupling $D_{\lambda,11}$ as functions of the DM mass $m_{\chi_b}$ in the single flavor freeze-out scenario. The mass hierarchy $m_{\chi_b}<m_{\chi_{d,s}}$ and the relic abundance constraint are imposed. The red points satisfy the bound from LUX, while the blue points satisfy the $\Delta F=2$ constraints. For the yellow points both LUX and $\Delta F =2$ constraints are imposed.}
\end{figure}

Figure \ref{fig:1f-s13-D11} shows the result of a parameter scan over $m_{\chi_b}$ and $\lambda$, with the relic abundance constraint for the single flavor freeze-out scenario imposed. The red points satisfy the bound from LUX, while the blue points satisfy the $\Delta F=2$ constraints. The yellow points fulfill both LUX and $\Delta F =2$ constraints. 

From the left panel, showing $s^\lambda_{13}$ as a function of $m_{\chi_b}$, we can see that both the LUX and the $\Delta F =2$ constraints require $s^\lambda_{13}$ to be small. In case of LUX this constraint arises from the necessary suppression of the DM-nucleon scattering at tree level. The $\Delta F =2$ constraints on the other hand require $s^\lambda_{13}$ to be small, as the mass splitting between $\chi_{d}$ and $\chi_b$ requires a deviation from the 13-degeneracy scenario $D_{\lambda,11}=D_{\lambda,33}$. This bound becomes stronger for small $m_{\chi_b}$ for the following reason: Small $m_{\chi_b}$ requires a large $D_{\lambda,33}$ from the relic abundance constraint, so that $|\eta|_\text{max}$ decreases with decreasing $m_{\chi_b}$. In turn a larger splitting between $D_{\lambda,11}$ and $D_{\lambda,33}$ is required in order to generate the $\gsim 10\%$ mass splitting, implying smaller flavor mixing $s^\lambda_{13}$.
We further observe that once both the LUX and the $\Delta F=2$
constraints are taken into account,  the upper bound on
$s^\lambda_{13}$  gets stronger than the individual ones. This proves
a non-trivial interplay of the flavor and DM constraints and underlines the importance of a combined study.

In the right panel we show the allowed size of $D_{\lambda,11}$ as a function of $m_{\chi_b}$. The upper bound on $D_{\lambda,11}$ as a function of $m_{\chi_b}$ arises from the relic abundance constraint on $D_{\lambda,33}$ together with the hierarchy requirement $D_{\lambda,11}<D_{\lambda,33}$. For $m_{\chi_b}\gsim 100\gev$ the whole range of $D_{\lambda,11}$ up to the relic abundance bound is allowed. However for smaller $m_{\chi_b}$ the LUX constraint disfavors large values for $D_{\lambda,11}$, so that in combination with the $\Delta F=2$ constraints a sharp cutoff arises. This cutoff decreases with decreasing mass $m_{\chi_b}$.
Below $m_{\chi_b} \sim 60\gev$ also a lower bound on $D_{\lambda,11}$
arises for the parameter points that satisfy both the LUX and flavor
constraints -- interestingly this bound does not emerge if the
constraints are taken into account separately. This is another clear
sign of a non-trivial interplay of the flavor and DM constraints.

Analytically the bounds on $D_{\lambda,11}$ can be understood by having a closer look at the structure of the WIMP-nucleon cross-section in the mDMFV model. As already mentioned above, the $s$-channel $\phi$ exchange at tree level is proportional to $(D_{\lambda,33}s^\lambda_{13})^2$ and therefore places a strong constraint on $s^\lambda_{13}$. We are then left with the one-loop box and photon penguin contributions. 
The box amplitude is positive and proportional to 
\be\label{eq:box-dependence}
D_{\lambda,33}^2\cdot\left[ (D_{\lambda,11}c^\lambda_{12})^2+(D_{\lambda,22}s^\lambda_{12})^2\right]\,.
\ee
It can therefore be suppressed by choosing both terms in the bracket
small. This explains why the LUX constraint, which is strongest for
low DM masses, gives rise to an upper bound on $D_{\lambda,11}$. 

The size of the photon penguin is determined by $D_{\lambda,33}^2$ and therefore fixed by the relic abundance (there is a small dependence on $s^\lambda_{23}$, due to the difference in quark masses). Interestingly due to the log factor the photon penguin amplitude carries an overall minus sign. This relative sign between the penguin and box amplitudes leads to a cancellation of the two contributions, provided the expression in the bracket of \eqref{eq:box-dependence} is of the right size. In the absence of flavor constraints $D_{\lambda,11}$, $D_{\lambda,22}$ and $s^\lambda_{12}$ are independent parameters. Consequently no conclusion can be drawn on the value of only one of them, while letting the other two vary. Taking into account the flavor constraints the picture changes however. In that case $s^\lambda_{12}$ is allowed to be sizable only if $D_{\lambda,22}\simeq D_{\lambda,11}$. Eq.\ \eqref{eq:box-dependence} then  reduces to $D_{\lambda,33}^2\cdot D_{\lambda,11}^2$, and we can 
directly read off the requirement that $D_{\lambda,11}$  has to lie in
a specific range such that the cancellation between the penguin and box contributions can work. 

This mechanism provides a realization of xenophobic DM
\cite{Feng:2011vu,Feng:2013vod}. The box diagram contributes to DM
scattering off protons and neutrons, while the photon penguin is only
present for DM-proton scattering. Therefore the penguin-box
cancellation works only for a specific number of protons and neutrons
in the nucleus.

We note that a pattern of constraints analogous to the one observed in
figure \ref{fig:1f-s13-D11} arises for all the possible mass
hierarchies in the dark sector, i.\,e.\ it does not depend
qualitatively on whether a single flavor or multiple flavors are
present at freeze-out.

\begin{figure}[h!]
\centering
\begin{minipage}{7.5cm}
\includegraphics[width=\textwidth]{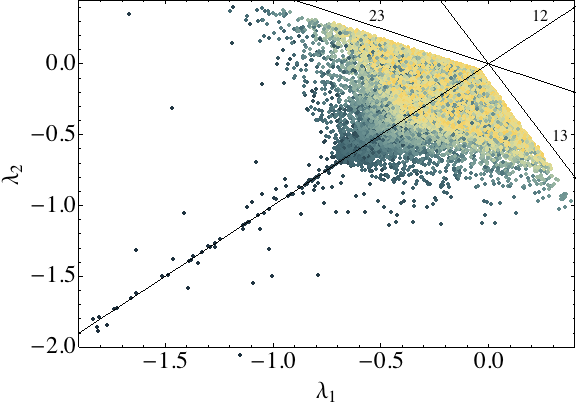}
\end{minipage}\hspace{1mm}
\begin{minipage}{2.2cm}
\includegraphics[width=\textwidth]{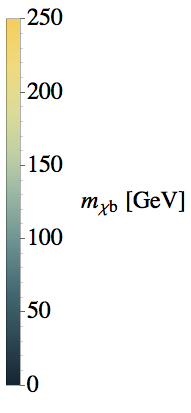}
\vspace*{.01cm}

\end{minipage}
\caption{\label{fig:1f-lambda1-lambda2}Allowed region in the
$\lambda_1$-$\lambda_2$-plane for the single flavor freeze-out
scenario, after imposing the relic abundance, LUX and flavor
constraints. The DM mass $m_{\chi_b}$ is indicated by the color, and the $ij$-degeneracy lines are sketched.}
\end{figure}

Next let us recover which of the flavor scenarios identified in section \ref{sec:DF2numerics} are realized in the single flavor freeze-out case. To this end we parameterize the matrix $D_\lambda$ in terms of the parameters $\lambda_{0,1,2}$, used in the flavor analysis, and defined in \eqref{eq:Dlambda}.
The distribution of parameter points consistent with the LUX and flavor constraints in the $\lambda_1$-$\lambda_2$-plane is shown in figure \ref{fig:1f-lambda1-lambda2}. We observe that the requirement $m_{\chi_b}<m_{\chi_{d,s}}$ and therefore $D_{\lambda,11},D_{\lambda,22}<D_{\lambda,33}$ restricts the allowed parameter space to the region $\lambda_2< -1/2\lambda_1 \wedge \lambda_2< -2\lambda_1$. Consequently only the 12 degeneracy scenario and the small mixing scenario are realized in this case.
The DM mass $m_{\chi_b}$ is indicated by the color, with the dark
points corresponding to light DM. We see that with smaller DM mass $m_{\chi_b}$ points move further away from the universality $\lambda_1 = \lambda_2=0$. This feature is a direct consequence of the upper bound on $D_{\lambda,11}$ for  $m_{\chi_b}\lsim 100\gev$, requiring a sizable splitting in $D_\lambda$.

\subsection{13-degeneracy}

Next we consider the scenario of quasi-degenerate first and third dark generations, i.\,e.\ $m_{\chi_b}\lsim m_{\chi_d}< m_{\chi_s}$. In this case we have $D_{\lambda,33}\gsim D_{\lambda,11} > D_{\lambda,22}$. Comparing this structure to the scenarios identified in the flavor pre-analysis we see that we are confined to the 13-degeneracy scenario, in which $s^\lambda_{12}$ and  $s^\lambda_{23}$ are small while  $s^\lambda_{13}$ is in principle allowed to be large. However recalling the results from the previous section we anticipate that once the LUX bound is imposed also the latter mixing angle is constrained to be small, in order to suppress the tree level contribution to the WIMP-nucleon scattering. This is indeed confirmed by our numerical analysis.

\begin{figure}[h!]
\centering
\includegraphics[width=.49\textwidth]{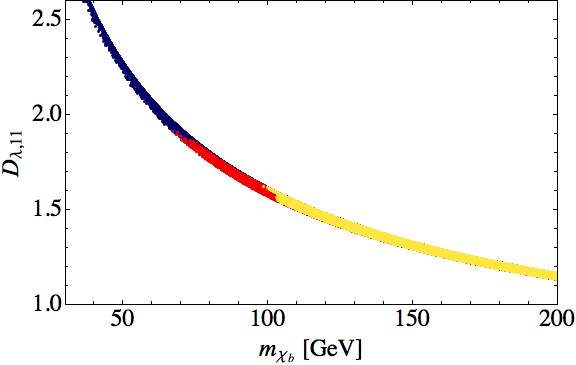}
\caption{\label{fig:2f-D11-13deg}First generation coupling
$D_{\lambda,11}$ as function of the DM mass $m_{\chi_b}$ in the 13-degeneracy scenario.
The mass hierarchy $m_{\chi_b}\lsim m_{\chi_{d}}<m_{\chi_{s}}$ and the relic abundance constraint are imposed. The red points satisfy the bound from LUX, while the blue points satisfy the $\Delta F=2$ constraints. For the yellow points both LUX and $\Delta F =2$ constraints are imposed.}
\end{figure}

In figure \ref{fig:2f-D11-13deg} we show the allowed values of the
first generation coupling $D_{\lambda,11}$ as a function of the DM mass $m_{\chi_b}$. We observe that due to the quasi-degeneracy of the first and third generation for a given $m_{\chi_b}$ only a small range of parameter space is allowed by the relic abundance constraint. Consequently the upper bound on $D_{\lambda,11}$ that arises again from the need to cut off the box contribution to the WIMP-nucleon scattering translates into a lower bound on $m_{\chi_b}$. If we neglect the flavor constraints and
take into account only the LUX bound, we find a lower bound $m_{\chi_b}\gsim 70\gev$. Taking into account both the LUX and flavor constraints simultaneously, the bound becomes considerably stronger, $m_{\chi_b}\gsim 100\gev$.
Again we stress the non-trivial interplay of flavor and DM constraints.

\begin{figure}[h!]
\centering
\begin{minipage}{7.5cm}
\includegraphics[width=\textwidth]{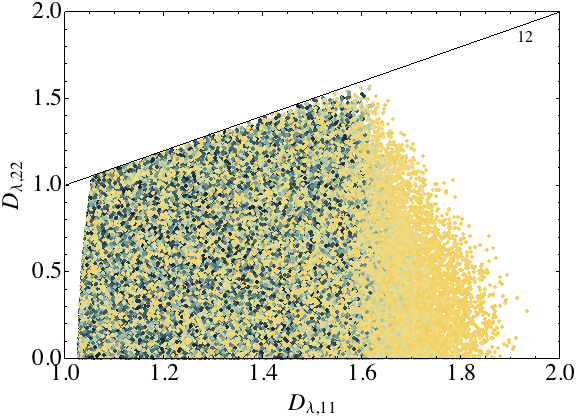}
\end{minipage}
\begin{minipage}{1.4cm}
\includegraphics[width=\textwidth]{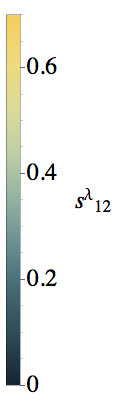}
\vspace*{.01cm}

\end{minipage}
\caption{\label{fig:2f-13deg-D11-D22-LUX}Allowed region for $D_{\lambda,11}$ and $D_{\lambda,22}$ in the 13-degeneracy scenario, after imposing both the relic abundance and LUX constraints, while the flavor constraints are not taken into account. The flavor mixing angle $s^\lambda_{12}$ is indicated by the color, and the 12-degeneracy line $D_{\lambda,11}=D_{\lambda,22}$ is sketched.}
\end{figure}

The interplay of flavor and DM constraints can be understood from
figure \ref{fig:2f-13deg-D11-D22-LUX}. We show the allowed parameter
range in the $D_{\lambda,11}$-$D_{\lambda,22}$-plane, applying only
the LUX bound, but not the flavor constraints. We observe that the
largest values for $D_{\lambda,11}$, corresponding to the smallest DM masses, are reached away from the 12-degeneracy line and require near-maximal mixing $s^\lambda_{12}$.
 This part of parameter space, while perfectly consistent with the DM constraints, is however ruled out by the stringent $\Delta F=2$ constraints that allow a large $s^\lambda_{12}$ only very close to the 12-degeneracy line.

\begin{figure}[h!]
\centering
\begin{minipage}{7.5cm}
\includegraphics[width=\textwidth]{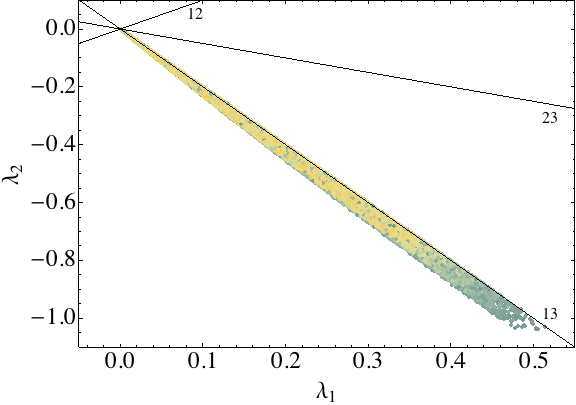}
\end{minipage}\hspace{1mm}
\begin{minipage}{2.2cm}
\includegraphics[width=\textwidth]{figures/mchi-color.png}
\vspace*{.01cm}

\end{minipage}
\caption{\label{fig:2f-13deg-lambda1-lambda2}
Allowed region in the $\lambda_1$-$\lambda_2$-plane for the
13-degeneracy scenario, after imposing the relic abundance, LUX and
flavor constraints. The DM mass $m_{\chi_b}$ is indicated by the color, and the $ij$-degeneracy lines are sketched.}
\end{figure}

In figure \ref{fig:2f-13deg-lambda1-lambda2} we show the allowed region of parameter space in the $\lambda_1$-$\lambda_2$-plane.
As expected all parameter points lie very close to the exact
13-degeneracy line $\lambda_2=-2\lambda_1$. With decreasing DM mass $m_{\chi_b}$ they move further away from the universality point $\lambda_1 = \lambda_2=0$. This is a consequence of the required splitting of the second generation $m_{\chi_s}$.

\subsection{23-degeneracy}

\begin{figure}[h!]
\centering
\begin{minipage}{7.5cm}
\includegraphics[width=\textwidth]{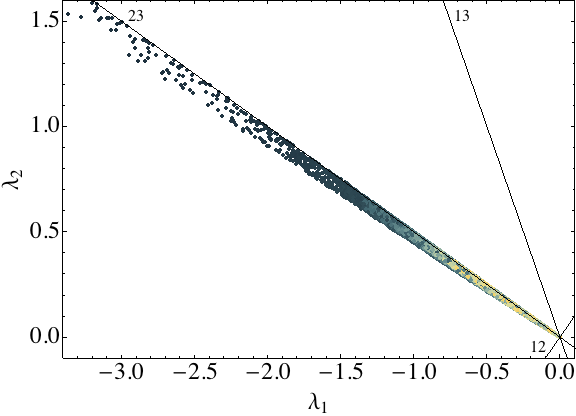}
\end{minipage}\hspace{1mm}
\begin{minipage}{2.2cm}
\includegraphics[width=\textwidth]{figures/mchi-color.png}
\vspace*{.01cm}

\end{minipage}
\caption{\label{fig:2f-23deg-lambda1-lambda2}
Allowed region in the $\lambda_1$-$\lambda_2$-plane for the
23-degeneracy scenario, after imposing the relic abundance, LUX and
flavor constraints. The DM mass $m_{\chi_b}$ is indicated by the color, and the $ij$-degeneracy lines are sketched.}
\end{figure}

The next scenario to study is the approximate 23-degeneracy, with
$m_{\chi_b}\lsim m_{\chi_s}< m_{\chi_d}$. In this case
$s^\lambda_{23}$ is arbitrary, with $s^\lambda_{12}$ is required to be
small by the $\Delta F =2 $ constraints, and both $\Delta F =2 $ and
LUX constraints demand $s^\lambda_{13}$ to be small. In figure
\ref{fig:2f-23deg-lambda1-lambda2} we show the allowed parameter space
for $\lambda_1$ and $\lambda_2$, which as expected is confined to be
very close to the 23-degeneracy line $\lambda_2=-1/2\lambda_1$. Again
small DM mass $m_{\chi_b}$ 
implies a sizable coupling non-universality $\lambda_{1,2}\ne 0$. In contrast to the 13-degeneracy in this case no lower bound on $m_{\chi_b}$ emerges.

\subsection{Quasi-degenerate dark sector}

We finally study the case of a quasi-degenerate dark sector with mass
splittings $<1\%$. Such a scenario can be achieved either by having a
quasi-universal coupling matrix $D_\lambda$ or by suppressing the DMFV
correction to the mass matrix by the smallness of the parameter
$\eta$. Note however that the correction is unavoidably generated at
one loop level, so $|\eta|\lsim 10^{-2}$ would involve fine-tuning. As
in the previous cases we scan over the allowed parameter space for
$\lambda$, assuming $b$-flavored DM and requiring the relic
abundance to be set by thermal freeze-out. Since the three $\chi$
flavors are quasi-degenerate, all of them are present at 
freeze-out.

\begin{figure}[ht]
\centering
\begin{minipage}{7.5cm}
\includegraphics[width=\textwidth]{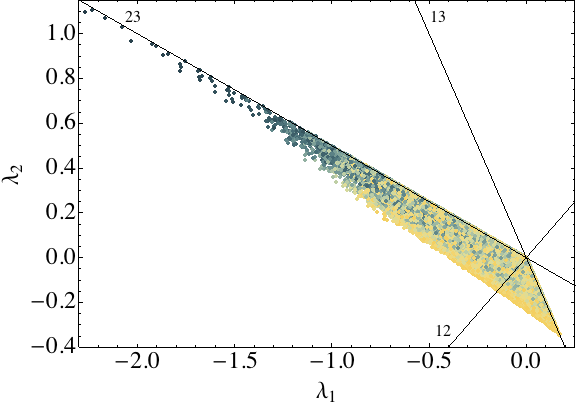}
\end{minipage}\hspace{1mm}
\begin{minipage}{2.2cm}
\includegraphics[width=\textwidth]{figures/mchi-color.png}
\vspace*{.01cm}

\end{minipage}
\caption{\label{fig:3f-lambda1-lambda2}Allowed region in the
$\lambda_1$-$\lambda_2$-plane for the quasi-degeneracy scenario, after
imposing the relic abundance, LUX and flavor constraints. The DM mass $m_{\chi_b}$ is indicated by the color, and the $ij$-degeneracy lines are sketched.}
\end{figure}

In figure \ref{fig:3f-lambda1-lambda2} we show the allowed range of
parameters in the $\lambda_1$-$\lambda_2$-plane. The DM mass
$m_{\chi_b}$ is varied and depicted by the color. We observe that for
large DM masses $m_{\chi_b}\gsim 100\gev$ the allowed range of
parameters is indeed close to the universality scenario
$\lambda_{1,2}=0$. Some deviations are however allowed, as the
non-universal corrections can be suppressed by a small expansion
parameter $\eta$. For small DM masses however a significant deviation $\lambda_1\lsim -1$ from the universality scenario is necessary and therefore $\eta$ has to be loop suppressed. A sizable negative $\lambda_1$, together with a near 23-degeneracy, suppresses the coupling $D_{\lambda,11}$ with respect to the other $D_\lambda$ components. Consequently the relic abundance constraint is brought in accordance with the bound from LUX.

\section{Collider phenomenology}\label{sec:collider}
In this section we present the collider constraints {on the mDMFV model}. We expect
significant constraints from the LHC since our model contains new
colored states, as well as large coupling to quarks. 
Our framework predicts a number of new particles within the reach of the LHC: 
\begin{itemize}
\item
The scalar mediator $\phi$ which carries QCD charge and therefore has
a large production cross-section in proton-proton collisions. 
\item
The DM particle $\chi_b$ and the heavier flavors $\chi_{d,s}$, which
are Dirac fermions and singlets under the SM gauge group. 
\end{itemize}

This spectrum has some similarities to simplified models of squarks
and neutralinos in the MSSM. The scalar mediator $\phi$
is similar to a right-handed down-type squark, except that it does
not carry flavor. It couples to all three quark flavors $d,s,b$ with
strengths given by the elements of the coupling matrix $\lambda$.
The flavor is carried by $\chi$, leading to three
somewhat degenerate states, in contrast to a neutralino LSP (lightest
supersymmetric particle). Also note
that in the simplest versions of supersymmetry neutralinos are
Majorana fermions, whereas $\chi$ is a Dirac fermion.

A full treatment of collider constraints is beyond the scope of this work.
We use existing analyses and translate their bounds into limits on our
model. There are some qualitatively new features beyond the
simplified MSSM model. First is the presence of large DM
trilinear couplings with quarks. In the {MSSM, supersymmetry} constrains this
coupling to be of the electroweak strength, making the production of
squarks insensitive to the coupling with DM. In contrast, the
large couplings required for relic abundance in the {mDMFV model}
contribute non-negligibly to the production. This also makes the
production sensitive to the flavor structure of the DM-quark
couplings.

While large flavor angles in the mDMFV model are an interesting
possibility, they lead to a large number of free parameters, and 
we postpone a detailed analysis of this case for future
work. Here, we focus on the cases where the mixing angles are
small. This has implications for both production and decay of $\phi$
and $\chi$.

The main constraints on the mDMFV model come from searches for
monojets and dijets in conjunction with missing transverse energy. We
now estimate what constraints these searches put on our model. We also
point out some unexplored novel signatures that arise in this framework.

\subsection{Multijet with missing energy searches}

Let us begin with the constraints from multijet + MET searches. 
The relevant analyses that we
consider for this case are the 19.5 fb$^{-1}$ CMS analysis for direct
sbottom production~\cite{CMS-PAS-SUS-13-018} and multijet + MET
analysis for squark-pair
production~\cite{Chatrchyan:2014lfa}. The colored particle $\phi$ is analogous to a
squark in the SUSY simplified models considered in these analyses, 
but with some important differences which we highlight below.

A dominant mode of $\phi$
pair production is through QCD interactions, in analogy with pair
production of a single squark {flavor}, in the limit of decoupled gluino mass.
There are two additional contributions that we need to consider.
There is a sizeable contribution from the $t$-channel process mediated
by $\chi_d$ due to the valence down quark pdf.  Depending upon the
value of the coupling $D_{\lambda, 11}$, this can be an
$\mathcal{O}$(1) fraction of the total production rate. The
corresponding contribution from the $D_{\lambda,22}$ is suppressed by
the sea-quark $s$-quark pdf and does not contribute significantly.

There are also contributions from an off-shell quark $(gg\to d d^* \to
d \chi_d \phi)$ which yield the same dijet + MET final state. The total
production cross section for this process can be much higher than the
QCD/$t$-channel production. However, this contribution always involves
one jet arising from a gluon splitting, which typically carries a very
small $p_T$. Thus, most events in this channel do not pass the
selection cuts for dijet analysis.

As is familiar from squark pair-production, NLO QCD effects are
significant. Since there is no NLO calculation available for our
model, we present limits for two extreme cases: one is assuming the
leading order production cross section, which provides a weak bound.
The other limit is to take the squark production
K-factor~\cite{Kramer:2012bx} , and apply
it to the LO cross section obtained above. Note that for the case
$D_{\lambda, 11} =0$, the production cross section is given directly by
the squark pair production cross section.

Another important difference from the sbottom case is the decay of the
mediator $\phi$. In presence of small mixing, squarks decay into a
flavor-specific jet and the LSP, whereas $\phi$ can decay to each of
the quark flavors $d$, $s$ and $b$. This has the effect of weakening
the limits relative to the sbottom search for sizeable branching
fraction to the first and second generation. {A similar effect has
been observed for the case of stop scharm mixing in
\cite{Blanke:2013uia,Agrawal:2013kha}, where the bounds on the stop
mass get weakened due to the light jets in the final state.}

In the CMS sbottom search~\cite{CMS-PAS-SUS-13-018}, separate search regions are defined
for one and two $b$-tags. For the SUSY simplified model, the two $b$-tag search
region is the most sensitive for almost all of the parameter space. To
translate limits from this analysis to the mDMFV case, we of course
need to account for the branching fractions of $\phi$. The scaling is
different for the one vs. two $b$-tag region, and it is possible that for
some part of parameter space the one $b$-tag region is more sensitive. For
the current study we scale the cross section assuming the dominance of
the two $b$-tag signal region.

There is no $b$-tag veto employed in the CMS squark search.
Consequently, we can directly use the cross section limits obtained in
that case. These limits are seen to be stronger than the sbottom
limits for lower $m_\phi$ values, or for higher values of
$D_{\lambda,11}$. The latter is easily explained by the fact that
in sbottom searches the increased
production cross section for larger $D_{\lambda,11}$ is offset by a
smaller branching ratio into b-quarks.

\begin{figure}[tp]
  \begin{center}
    \includegraphics[width=0.45\textwidth]{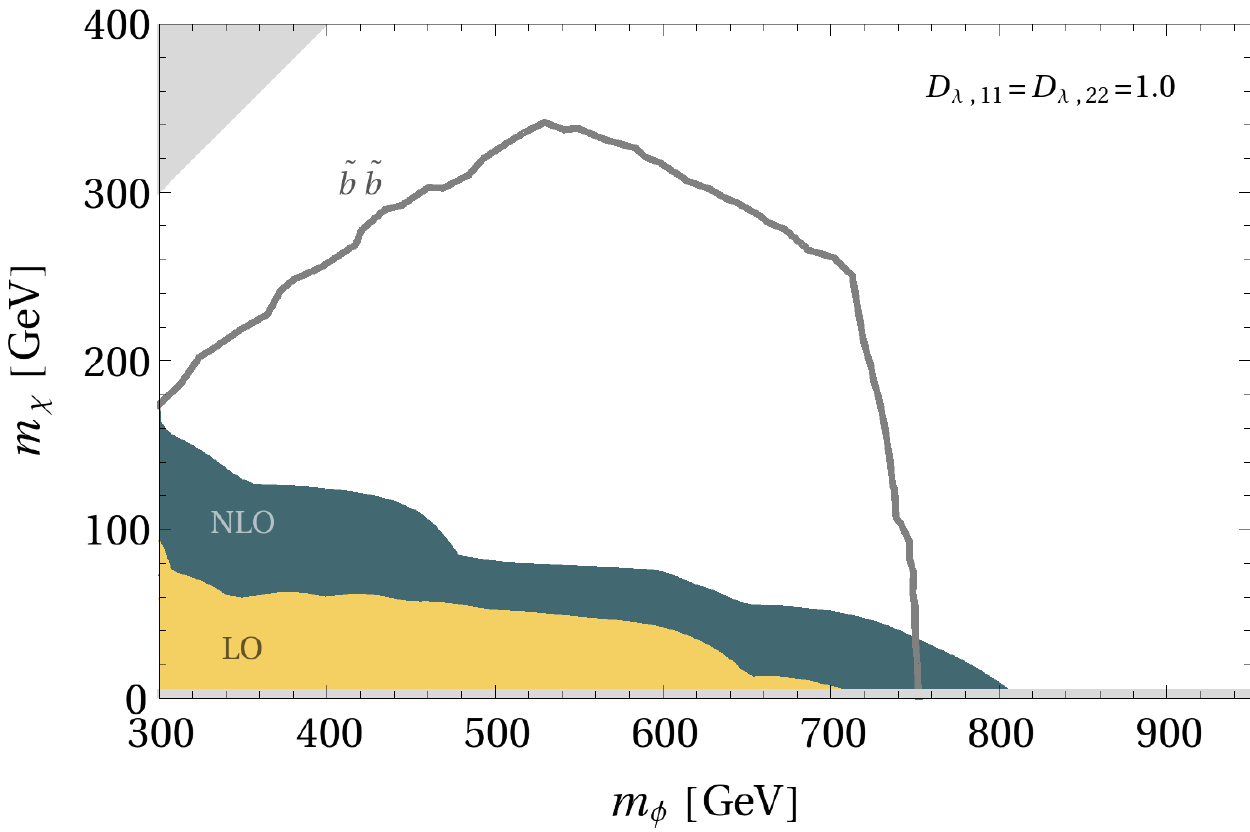}
    \quad
    \includegraphics[width=0.45\textwidth]{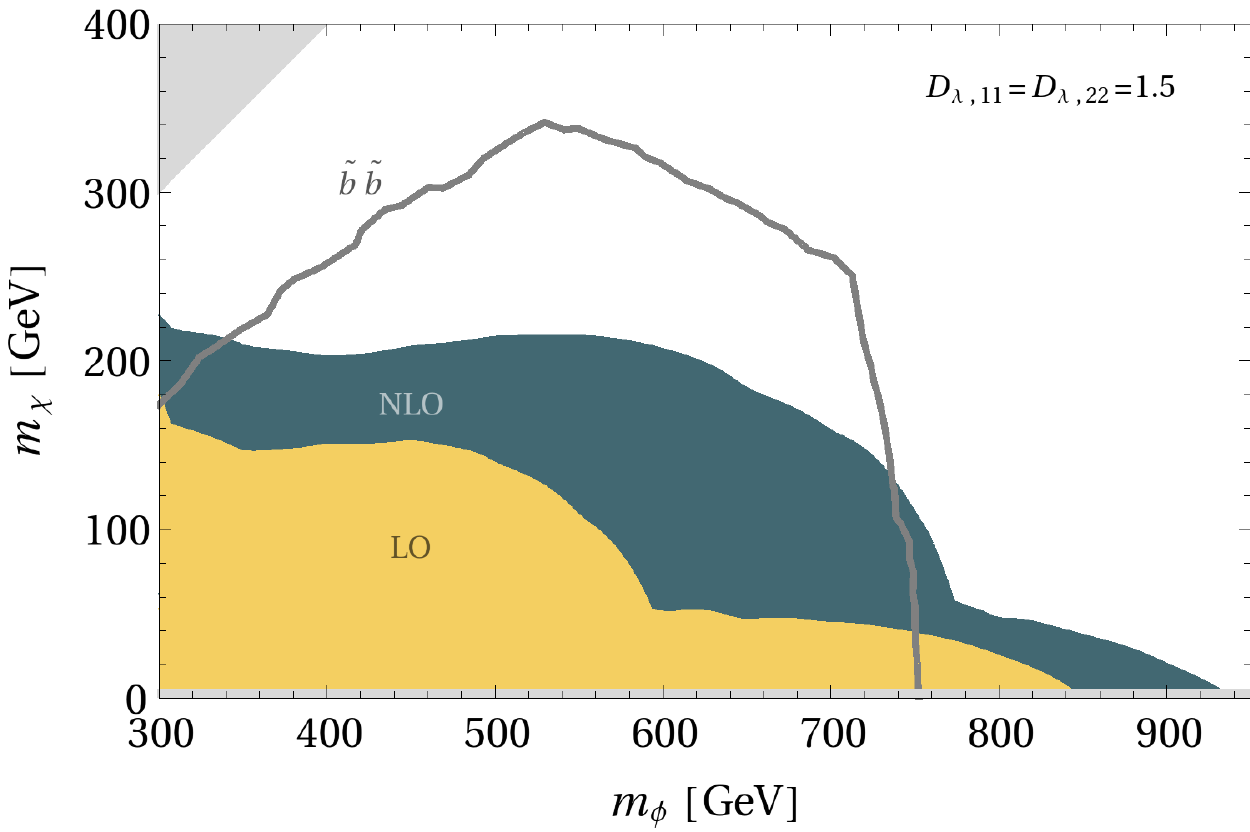}
  \end{center}
  \caption{Limits on the mDMFV model from the CMS 19.5
  fb$^{-1}$ sbottom~\cite{CMS-PAS-SUS-13-018} and
  squark~\cite{Chatrchyan:2014lfa} search. We show for comparison the
  CMS $\tilde{b}\tilde{b}$ limit (grey curve), which also directly
  applies to the case
  $D_{\lambda,11}=D_{\lambda,22}=0$. Shaded regions are excluded for
  particular choices of $D_{\lambda,11}$ and $D_{\lambda,22}$ as shown on
  the plot. For the NLO limit, we estimate
  the cross section by multiplying 
  LO prediction with the NLO K-factor from sbottom pair
  production. The DM coupling to $b$-quark, $D_{\lambda,33}$
  is fixed everywhere by the corresponding relic abundance
  constraint, and all mixing angles are set to zero for simplicity.}
  \label{fig:sbottoms1}
\end{figure}

We present our results in the $m_{\phi}$\,--\,$m_\chi$ mass plane in figure
\ref{fig:sbottoms1}. We choose three benchmark coupling values,
$D_{\lambda,11}=D_{\lambda,22}=\{0,1,1.5\}$ to
demonstrate the qualitative nature of the limits. The coupling
$D_{\lambda,33}$ is fixed to yield the correct relic abundance for each
point on the plane, assuming a single flavor freeze-out. We show the limits
from both the CMS 19.5 fb$^{-1}$ sbottom~\cite{CMS-PAS-SUS-13-018} and
squark~\cite{Chatrchyan:2014lfa} searches. The
NLO estimate is obtained from the LO cross section multiplied by
K-factors from sbottom production. Larger $m_\chi$ correspond to a
smaller $D_{\lambda, 33}$, and hence to a smaller $\phi\to b\chi_b$
branching ratio for fixed $D_{\lambda,11}$ and $D_{\lambda,22}$. This
results in a much weaker limit for higher values of $m_\chi$ in figure
\ref{fig:sbottoms1}.

\begin{figure}[tp]
  \begin{center}
    \includegraphics[width=0.45\textwidth]{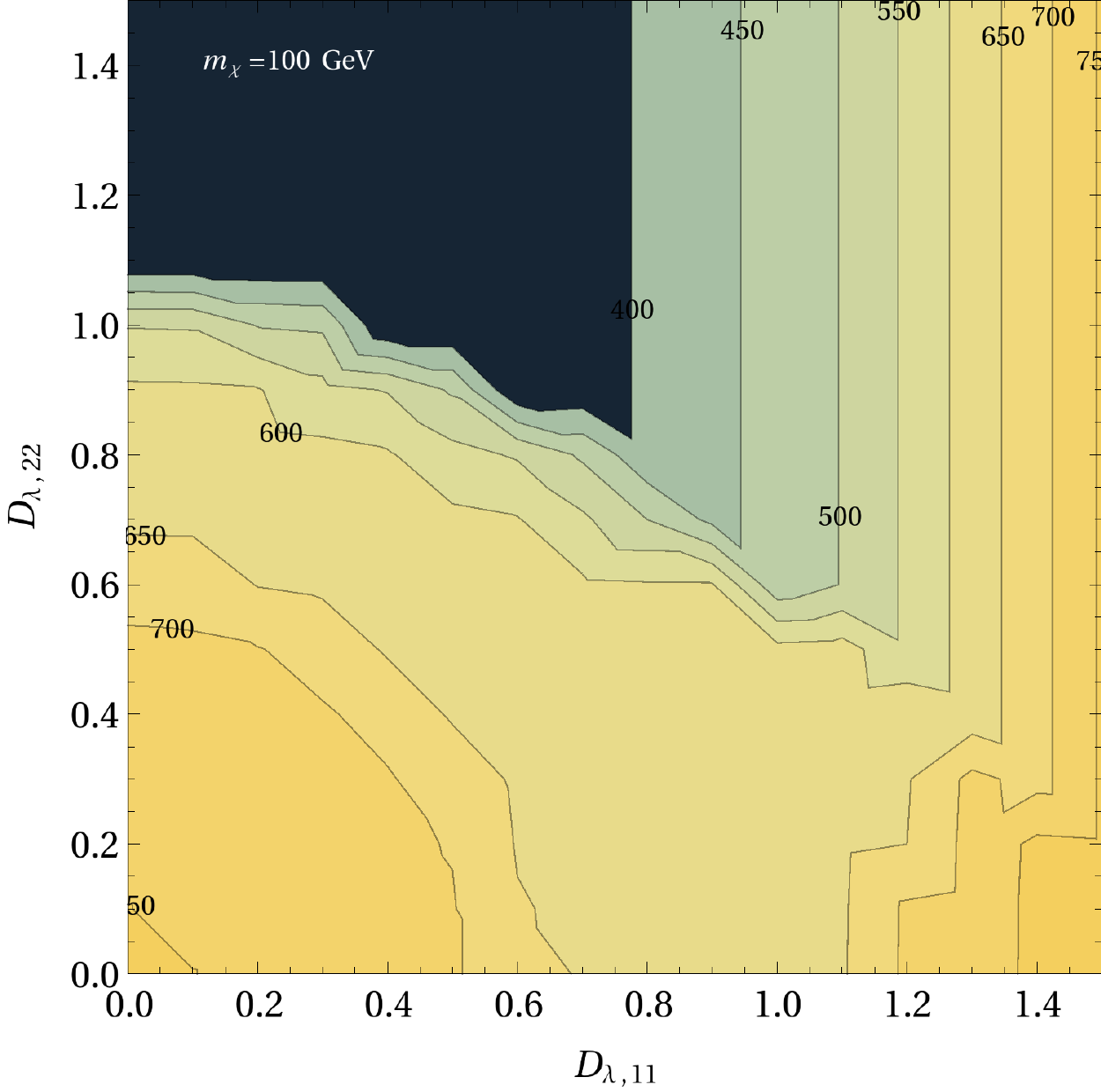}
  \end{center}
  \caption{Exclusion contours for $m_\phi$ {(in GeV)} for various values of
  $D_{\lambda,11}$ and $D_{\lambda,22}$ from the CMS
  sbottom~\cite{CMS-PAS-SUS-13-018} and
  squark~\cite{Chatrchyan:2014lfa} searches, obtained for the NLO
  estimate. $D_{\lambda,33}$ is fixed by the
  relic abundance constraint.}
  \label{fig:sbottoms2}
\end{figure}

In figure~\ref{fig:sbottoms2}, we present limits on $m_\phi$ for
different values of $D_{\lambda,11}$ and $D_{\lambda,22}$,  again with
$D_{\lambda,33}$ fixed by the relic abundance constraint. We choose a
representative DM mass of $m_\chi = 100$ GeV. This plot shows
the difference in sensitivity to the two different couplings -- the
constraint is much more severe for larger values of $D_{\lambda,11}$
than for corresponding values of $D_{\lambda,22}$ due to the enhanced
production cross section in the former case.

\subsection{Monojet searches} 

Another important constraint on the mDMFV parameter space is placed by
the searches for monojets and large missing $E_T$. There are two
interesting regimes for the monojet searches, the effective field
theory (EFT) regime {with a heavy mediator} and the 
compressed spectrum regime.

For large mediator masses, the dominant signal arises from
pair-production of DM along with a hard initial state
radiation (ISR) jet. This can be
crudely modeled by an EFT, which has the advantage of being largely
model independent. Limits are presented as lower bounds on $\Lambda$,
the scale associated with the coefficient of the higher dimensional
operator~\cite{CMS-PAS-EXO-12-048}. However, exclusions using this
technique have to be interpreted
with care. In principle, the validity of the EFT fails for $\Lambda
\lesssim p_T$, the $p_T$ cut employed in the searches. Thus,
technically we can only exclude a band in $\Lambda$ self-consistently
where the EFT is expected to be valid.  CMS and ATLAS have put limits
on various DM-quark EFT operators.  The flavor specific
operators that arise in our case have not been considered.  However,
we can estimate these limits by matching cross sections in our model
with those from the DM EFT used by CMS/ATLAS. 

In this case it is important to note that diagrams with an initial
state $d$ quark are enhanced by the valence quark pdf. This process is
therefore directly sensitive to the coupling $D_{\lambda,11}$,
suppressing that coupling  would help to effectively evade the monojet
constraints. In the DM mass regime we consider, $10\gev<m_\chi<250\gev$, the limits do not vary significantly. We show two
benchmarks $m_\chi=10$ GeV and $m_\chi = 210$ GeV as examples in
figure \ref{fig:monojetscoup}.

\begin{figure}[tp]
  \begin{center}
    \includegraphics[width=0.45\textwidth]{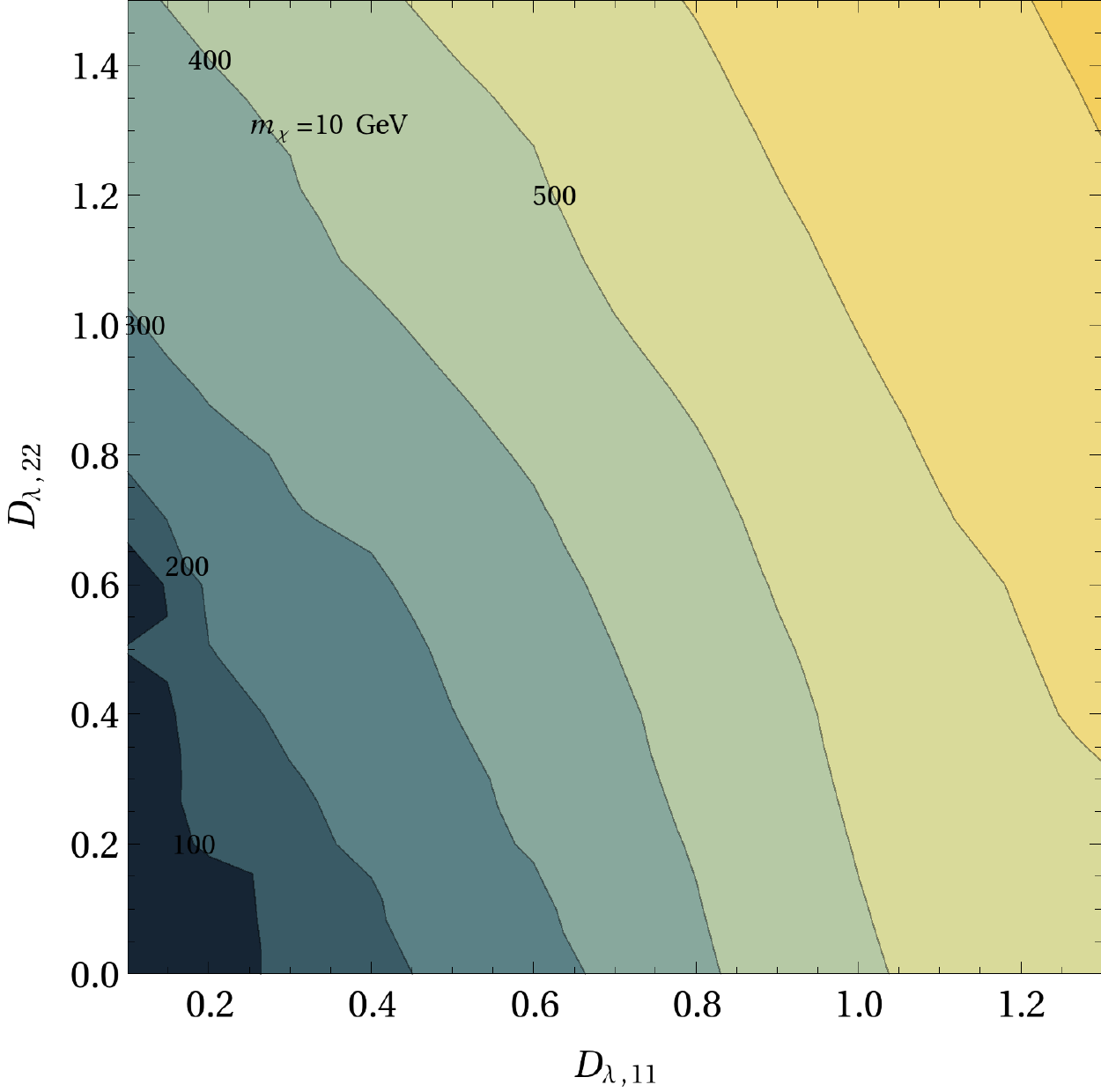}
    \qquad
    \includegraphics[width=0.45\textwidth]{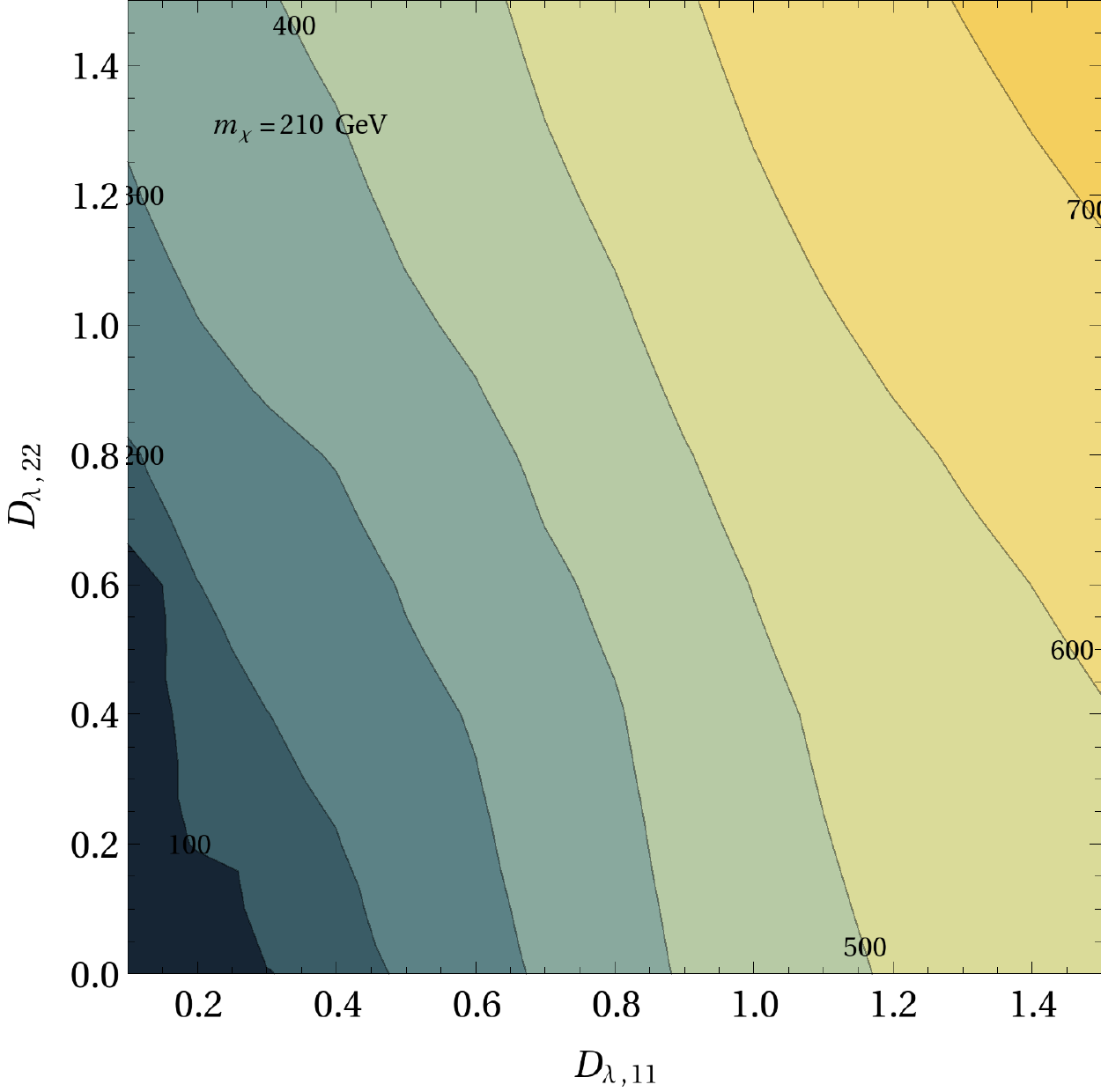}
  \end{center}
  \caption{Limits on the mDMFV model in terms of mass of $\phi$ (in
  GeV) from the CMS monojet
  search~\cite{CMS-PAS-EXO-12-048}. Limits are obtained by matching
  production cross sections in our model with those in the DMEFT used
  by CMS.  }
  \label{fig:monojetscoup}
\end{figure}

The second regime in which the monojets provide a constraint
complementary to the multijet searches is in the compressed region,
{$m_\phi \sim m_\chi $}. In this region, the jets from $\phi$ decays
are typically too soft to appear in multijet analyses. With a hard ISR
jet, the signal resembles the DM monojet signal. The
branching ratios of $\phi$ to different quark flavors are irrelevant
in this case, since these decay products are typically too soft to
pass the selection cuts. We show the limits obtained from recasting
the CMS search for stops decaying to charm and a neutralino
\cite{CMS-PAS-SUS-13-009} in figure
\ref{fig:monojets}.

\begin{figure}[tp]
  \begin{center}
    \includegraphics[width=0.45\textwidth]{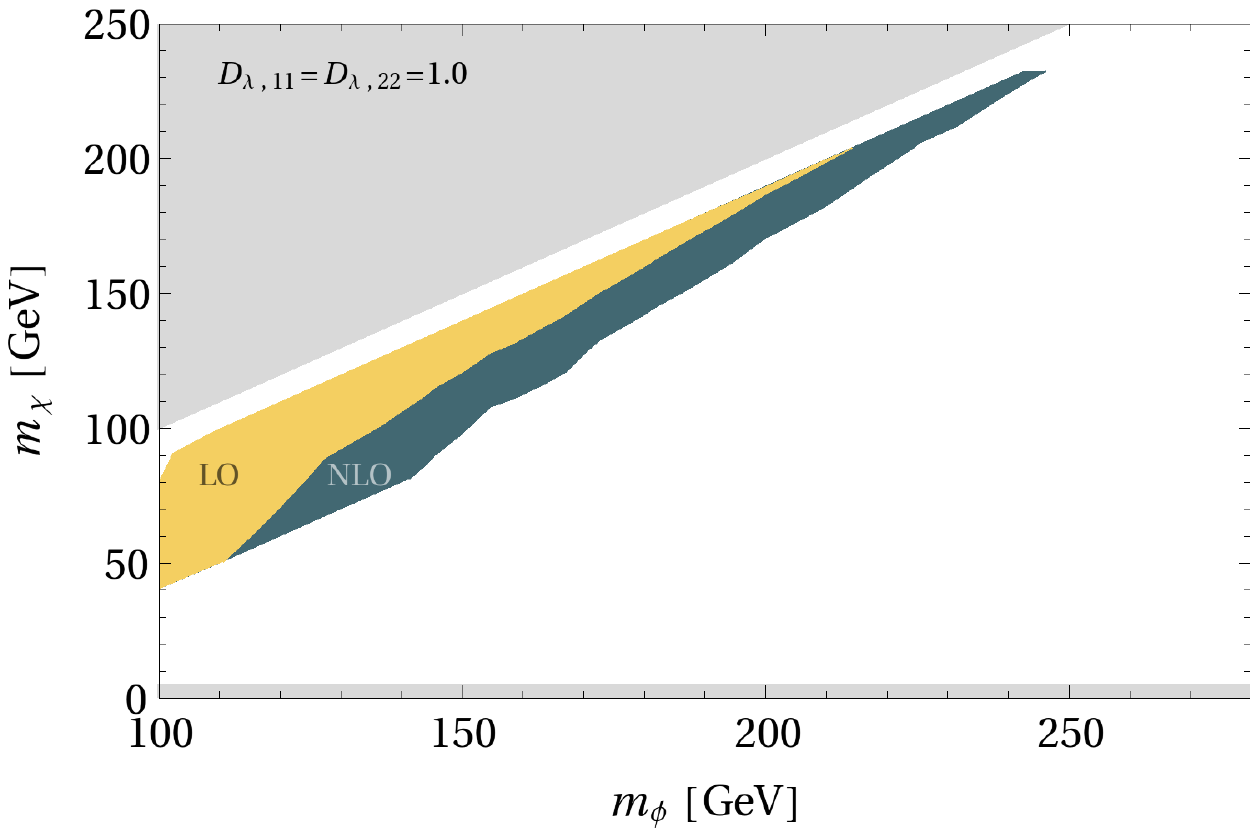}
    \includegraphics[width=0.45\textwidth]{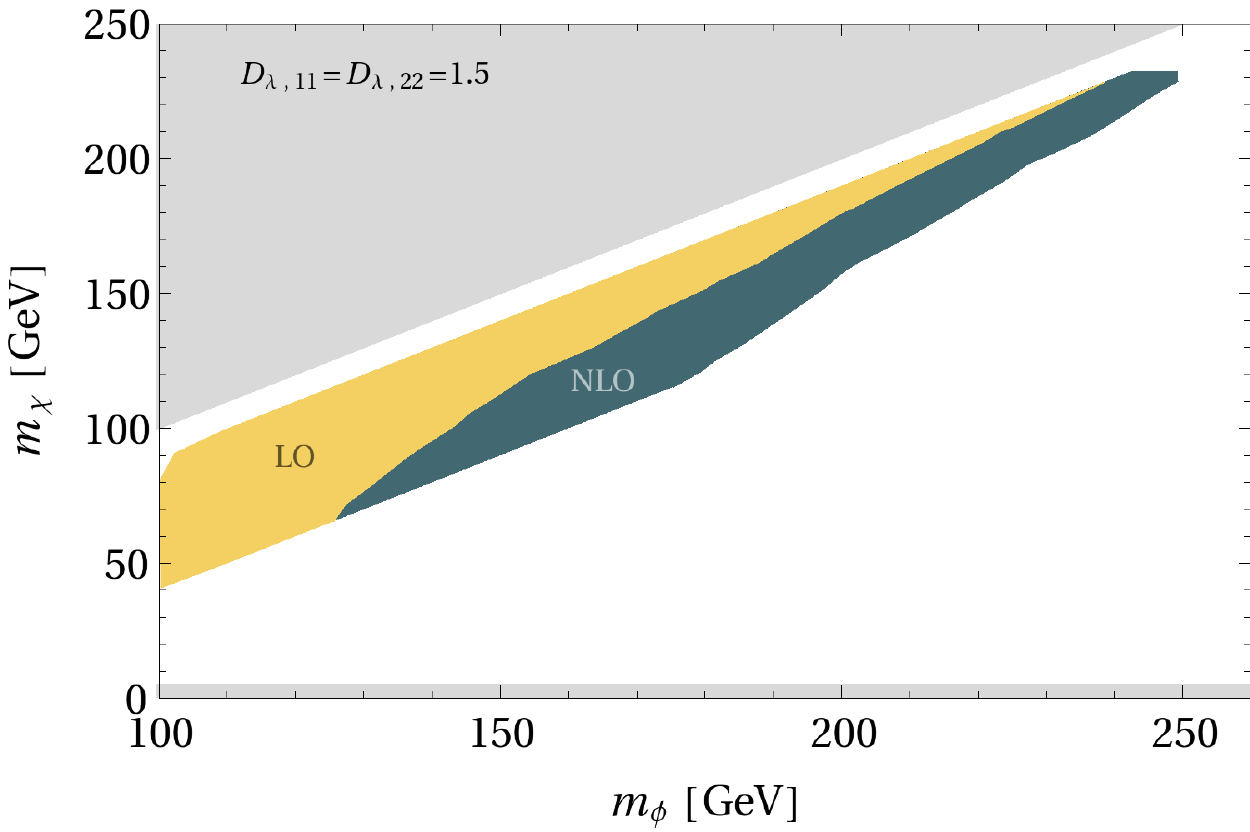}
  \end{center}
  \caption{Limits on the mDMFV model from the CMS 19.7
  fb$^{-1}$ search~\cite{CMS-PAS-SUS-13-009} for stops decaying to a
  charm and a neutralino,
  using the monojet + MET final state. 
  The NLO cross section is estimated
  from the LO prediction with the NLO K-factor from sbottom pair
  production. The triangular grey regions are parts of parameter space
  where $\phi$ is lighter than the dark matter particles and hence
  cannot decay into them.
  }
  \label{fig:monojets}
\end{figure}

Since the monojet analyses typically veto additional jets, they lose
sensitivity as the mass splitting between $\phi$ and $\chi$ increases.
It has been pointed out that the razor analysis can be powerful in
this region of parameter space~\cite{Fox:2012ee,Agrawal:2013kha,
Chatrchyan:2014goa}. It is an interesting future direction to analyze
the constraints on this framework from razor.

A more quantitative analysis of the constraints on the masses of
$\phi$ and $\chi_i$ would require a dedicated recasting of the
experimental analyses which is beyond the scope of the
present paper.

\subsection{Distinctive signatures}
We see that a number of current searches are sensitive to this model.
However, the presence of multiple flavors of DM have
potentially novel signatures which could distinguish these models with
other models of DM and colored mediators.

In the decay of $\phi$ all three DM flavors $\chi_i$ will be
produced. While the lightest state $\chi_b$ is stable and simply
escapes detection, the heavier states will decay to the lightest one
through the decay channels discussed in appendix \ref{sec:decayheavy}.

In the case when the mass splitting between the $\chi_i$ are small,
the decay products of the heavier states (jets, leptons or photons)
can be too soft by themselves for the current searches to be
sensitive. However, it is an interesting possibility to look for these
relatively soft objects in conjunction with high $p_T$ jets. In the
compressed regime, the heavier flavors can also have displaced decays,
or decays in the detector.

When the mass splitting is larger, we expect a multijet signal,
including heavy-flavor jets in addition to missing transverse
momentum. Observation of these cascade decays would help in
disentangling our model from other models of DM with
colored mediators.

\section{Conclusions}\label{sec:conclusions}

In this paper we presented a simplified model of flavored DM. The fermionic DM $\chi$ carries flavor and couples to right-handed down
quarks transmitted by a colored scalar mediator $\phi$.
We introduced the Dark Minimal Flavor Violation (DMFV) framework where the extended flavor symmetry $U(3)_q \times U(3)_u \times U(3)_d \times U(3)_\chi$ is broken only by the SM Yukawa couplings $Y_u$, $Y_d$ and the DM-quark coupling~$\lambda$.
In general the flavor violating effects
generated by $\lambda$ can be large. Hence DMFV, while being conceptually similar to MFV, is
phenomenologically very different from the latter scenario. 

The DMFV
assumption has the following implications that are very useful from a
model building perspective:
\bi
\item
If DMFV is exact, an unbroken $\mathbb{Z}_3$ subgroup stabilizes the DM. It also prevents the new particles from being singly produced at the LHC.
\item
The DMFV hypothesis allows a reduction in the number of free parameters. The matrix $\lambda$ can be parametrized in terms of a unitary matrix $U_\lambda$ and a positive and diagonal matrix $D_\lambda$.
\ei

We performed a detailed phenomenological analysis of the minimal DMFV
(mDMFV) model, focusing at first on the available constraints from flavor changing neutral current processes. 
Our findings can be summarized as follows:
\bi
\item Meson antimeson mixing observables put strong constraints on the structure of the coupling matrix $\lambda$. Yet it is possible to satisfy all existing $\Delta F=2$ constraints simultaneously if $\lambda$ obeys one of the following scenarios:
\begin{enumerate}
\item Universality scenario -- The diagonal coupling matrix $D_\lambda$ is nearly universal and large mixing angles in $U_\lambda$ are allowed.
\item Small mixing scenario  -- The mixing matrix $U_\lambda$ is close to the unit matrix, while $D_\lambda$ is arbitrary.
\item $ij$-degeneracy scenarios ($ij=12,13,23$) -- Two elements of $D_\lambda$ are almost equal, $D_{\lambda,ii}\simeq D_{\lambda,jj}$, and only the corresponding mixing angle $s^\lambda_{ij}$ is allowed to be large.
\end{enumerate}
Flavor violating effects are then efficiently suppressed by an almost diagonal matrix $\lambda \lambda^\dagger$. Our scenarios can straightforwardly be implemented in future studies. Note that their validity extends beyond the minimal model.
\item 
The mDMFV model does not yield any new contribution to rare
(semi-)leptonic $K$ and $B$ decays.\footnote{A non-SM experimental
signature for decays with neutrinos in the final state may arise if
the DM particle is light enough that the decay into a
$\chi\bar\chi$ final state is kinematically accessible
\cite{Kamenik:2011vy}.} Also the strongly constrained $B\to X_s\gamma$
decay does not receive any significant new contribution. Similarly 
new contributions to electroweak precision observables and
electric dipole moments are small in mDMFV.
\ei

In the second step the constraints from DM phenomenology are
included. We restricted our attention to the case of $b$-flavored DM
 which is phenomenologically most appealing. We also assumed
that the observed DM relic density arises from thermal
freeze-out. Our results are:
\bi
\item
The tree level contribution to DM-nucleon scattering, strongly
constrained by the recent LUX data, needs to be suppressed by a small
mixing angle $s^\lambda_{13}$.
\item 
Further contributions to DM-nucleon scattering arise at the one-loop
level from a box diagram and a photon penguin. These contributions
become relevant for DM masses $m_{\chi_b}\lsim 100\gev$.  The
penguin and box contributions carry a different overall sign, and
hence destructively interfere if they are comparable in magnitude. The
LUX constraint requires a partial cancellation, leading to both an
upper and a lower bound on the size of $D_{\lambda,11}$.
\item
We observe a non-trivial interplay of flavor and DM constraints, such
that the combined constraint on the parameter
space of $\lambda$ is an interesting overlap of the
individual ones.
\item
We studied several mass spectra implying different freeze-out conditions.
  In all cases the penguin box cancellation implies a sizable
  splitting $D_{\lambda,11}<D_{\lambda,33}$ for DM masses
  $m_{\chi_b}\lsim 100\gev$. This rules out the 13-degeneracy scenario
  below this scale.
\ei

Stringent constraints on the mDMFV model also arise from direct
searches at the LHC. We estimate the bounds arising from $b$-jets
$+\EmissT$ and monojet searches and point out possible new signatures.
We find:
\bi
\item 
Searches for direct sbottom quark production and for multiple jets plus missing energy
constrain the mDMFV parameter space up to $m_\phi\simeq 800-900\gev$.  
\item
Monojet searches are complementary to the searches for jets $+\EmissT$ and competitive in
two regions of parameter space, namely for large mediator masses or a compressed spectrum.
\item
The decay of the heavier $\chi$ flavors leads to additional jets or to
soft photons, which could be looked for in conjunction with the
monojet or dijet signals discussed above. 
\ei

Our pioneering study of flavored DM with a generic flavor
violating coupling matrix has revealed a lot of interesting phenomenology
and showed that it is worthwhile to consider DM models beyond MFV.

\subsection*{Acknowledgements}
PA and KG acknowledge partial support by the National Science Foundation under Grant No. PHYS-1066293 and the hospitality of the Aspen Center for Physics.
KG was supported by the
Deutsche Forschungsgemeinschaft (DFG), grant number GE 2541/1-1.
Fermilab is
operated by Fermi Research Alliance, LLC under Contract No. DE-AC02-07CH11359 with the United States
Department of Energy.

\begin{appendix}

\section{Classification of DMFV symmetry structures}\label{app:DMFV-classification}

In this appendix we classify the different possible versions of DMFV.
A common feature is that the SM global flavor symmetry is
extended by a $U(3)_\chi$ in the dark sector. In addition to the SM
Yukawa couplings there is only a single interaction $\lambda$ that
breaks the flavor symmetry. In principle it is straightforward to
implement DMFV in the lepton sector, but we restrict ourselves to the
quark sector in what follows. The global flavor symmetry is then given
by
\be
U(3)_q \times U(3)_u \times U(3)_d \times U(3)_\chi\,.
\ee
Throughout this paper we assumed that the DM $\chi$ is a  Dirac fermion and the mediator $\phi$ is a scalar, yet other Lorentz structures are also possible. We will not explore this option further.

In our analysis we considered the coupling of DM to the right-handed down type quarks by the structure
\be
\mathcal{L}\ni -\lambda \phi \bar d_R \chi_L\,.
\ee
In this case the coupling $\lambda$ induces the flavor symmetry breaking
\be
U(3)_d \times U(3)_\chi \to U(1)_{d\oplus\chi}\,.
\ee
We classify this symmetry breaking pattern as $\text{DMFV}_d$ in order to distinguish it from the following scenarios.

In $\text{DMFV}_u$ instead DM couples to the right-handed up quarks, 
\be
\mathcal{L} \ni -\lambda \phi \bar u_R \chi_L\,,
\ee
so that the symmetry breaking pattern is 
\be
U(3)_u \times U(3)_\chi \to U(1)_{u\oplus\chi}\,.
\ee
Due to the coupling to up-type quarks, the phenomenology of $\text{DMFV}_u$ is very different from the $\text{DMFV}_d$ case. We leave a detailed study for future work.

Last but not least it is also possible to couple DM to the left-handed quark doublets
\be
\mathcal{L} \ni -\lambda \phi \bar q_L \chi_R\,.
\ee
In this case either $\phi$ or $\chi$ has to transform as an $SU(2)_L$ doublet. In this so-called
{$\text{DMFV}_q$}
scenario, $\lambda$ yields the symmetry breaking
\be
U(3)_q \times U(3)_\chi \to U(1)_{q\oplus\chi}\,.
\ee
The phenomenology of this scenario is enriched by the fact that now DM couples to both up and down type quarks.

\section{\boldmath $\mathbbm{Z}_3$ as an unbroken subgroup of DMFV}\label{app:Z3}

In this appendix we present the proof for the presence of an unbroken
$\mathbbm{Z}_3$ subgroup emerging from the $SU(3)_\text{QCD} \times
U(3)_q\times U(3)_u\times U(3)_d \times U(3)_\chi$ symmetry in the
limit of exact DMFV, where $\chi$ transforms as a triplet under
$U(3)_\chi$. The argument is quite analogous to the one provided for the
MFV case in \cite{Batell:2011tc}. The unbroken $\mathbbm{Z}_3$
symmetry automatically forbids all operators inducing DM decay even at
the non-renormalizable level, {as long as they preserve DMFV}. It also prevents the new particles from
being singly produced at the LHC.  Note that the validity of the proof
does not depend on the Lorentz structure as it is based purely on
internal symmetry arguments.

Let us consider the operator
\be
\mathcal{O} \sim 
\chi \dots \bar \chi\dots \phi\dots \phi^\dagger\dots q_L\dots \bar q_L\dots u_R\dots \bar u_R \dots d_R\dots \bar d_R \dots G \dots \mathcal{S}\,,
\ee
where the dots represent an arbitrary number of insertions of the same fields, $G$ schematically denotes the gluon field or field strength and $S$ contains a combination of SM fields that are neutral under both QCD and the flavor symmetry. This is the most generic structure of interactions between the dark sector and SM particles. In the case $N_\chi=1, N_{\bar\chi}=N_\phi=N_{\phi^\dagger} =0$ it mediates DM decay.

This structure is invariant under QCD, if the number of $SU(3)_c$
triplet  minus the number of $SU(3)_c$ anti-triplets is a multiple of three:
\be
(N_\phi- N_{\phi^\dagger} + N_q + N_u + N_d - N_{\bar q} - N_{\bar u} - N_{\bar d}) \mod 3 = 0 \,.\label{eq:QCD}
\ee

Invariance of the operator $\mathcal{O}$ under the flavor symmetry can formally be achieved by treating the Yukawa couplings $Y_u$, $Y_d$ and $\lambda$ as spurion fields and adding a combination
\be
Y_u\dots Y_u^\dagger \dots Y_d\dots Y_d^\dagger\dots  \lambda\dots \lambda^\dagger\dots
\ee
to the operator $\mathcal{O}$.
Invariance under $U(3)_q$ then requires
\be
(N_q-N_{\bar q} + N_{Y_u} - N_{Y_u^\dagger} + N_{Y_d} - N_{Y_d^\dagger} )\mod 3=0 \,.\label{eq:U3q}
\ee
Similarly $U(3)_u$, $U(3)_d$ and $U(3)_\chi$ invariance give, respectively,
\begin{eqnarray}
  (N_u-N_{\bar u} - N_{Y_u} + N_{Y_u^\dagger} )\mod 3&=&0 \,,\\
  (N_d-N_{\bar d} - N_{Y_d} + N_{Y_d^\dagger} + N_\lambda - N_{\lambda^\dagger})\mod 3&=&0 \,,\\
  (N_\chi - N_{\bar \chi}- N_\lambda + N_{\lambda^\dagger})\mod 3&=&0 \,.\label{eq:U3chi}
\end{eqnarray}
Adding \eqref{eq:U3q}--\eqref{eq:U3chi} and subtracting \eqref{eq:QCD} we find
\be\label{eq:Z3}
(N_\chi - N_{\bar \chi} - N_\phi + N_{\phi^\dagger} )\mod 3 =0
\ee
as a necessary condition for the operator $\mathcal{O}$ to be invariant under QCD and the flavour symmetries. Eq. \eqref{eq:Z3} can be interpreted as a $\mathbb{Z}_3$ symmetry under which $\chi$ and $\phi$ carry the charges $e^{+2\pi i/3}$ and $e^{-2\pi i/3}$, respectively, while the SM fields carry the charge $+1$.

Consequently the combination of QCD and the DMFV assumption, together with our choice of representations forbid both $\chi$ and $\phi$ to decay into SM fields. Therefore {in the exact DMFV limit} the lightest state out of $\phi$ and the components of $\chi$ is stable without the need of imposing an additional symmetry.

\section{Relevant loop functions}
\label{app:functions}

In this appendix we collect the one loop functions relevant for our analysis.

The NP one loop function for $\Delta F=2$ processes in mDMFV reads
\be
F(x_i,x_j) = \frac{x_i^2 \log x_i}{(x_i-x_j)(1-x_i)^2} + \frac{x_j^2 \log x_j}{(x_j-x_i)(1-x_j)^2} + \frac{1}{(1-x_i)(1-x_j)}\,,
\ee
where $x_i=m_{\chi_i}^2/m_\phi^2$. In the limit of degenerate masses $m_\chi$ it reduces to
\be
F(x)= \frac{2x \log x}{(1-x)^3} + \frac{(1+x)}{(1-x)^2} 
\ee
with $x=m_\chi^2/m_\phi^2$.

The NP one loop contribution to the $b\to s\gamma$ transition is given by
\be
g(x) = \frac{1-5x-2x^2}{6(1-x)^3}-\frac{x^2\log x}{(1-x)^4}\,,\qquad x_i = \frac{m_{\chi_i}^2}{m_\phi^2}\,.
\ee

\section{Decay of heavier flavors}
\label{sec:decayheavy}
DM flavors pick up mass splitting either through threshold
corrections or through running. Consequently, the heavier $\chi$ flavors can decay into the lighter DM. The open decay modes
depend both on the size of mass splittings as well as on the structure
of the coupling matrix. As in the text, we focus on the case where
$\chi_b$ is the lightest flavor, and hence constitutes the DM
today. The relevant diagrams are shown in figure \ref{fig:decays}.

The dominant decay mode arises at tree-level, where the heavier DM
decays into a pair of quarks and a lighter DM.  This
decay mode is only open if the splitting between the DM
flavors is more than the lightest allowed mesons in the final state.
Since this is a flavor-conserving decay, it is expected to dominate
whenever allowed by phase space. Further, for couplings relevant for
us, this decay mode is prompt on the timescale of BBN.

For more compressed spectra, the quark/meson final states may not be
kinematically allowed. In this case, flavor-violating decays into a
photon or a pair of leptons via an off-shell photon will be present.
These can be estimated by using the effective coupling of DM
particles with a photon~\cite{Agrawal:2011ze,Kopp:2009et},
\begin{align}
  \mathcal{L}_{\rm eff} &= 
  \sum_{i=d,s,b}\frac{-\lambda_{ib}^* \lambda_{ih}e}{64\pi^2 m_\phi^2} \left[
  \left(
  \frac12
  +\frac23\log\left[\tfrac{m_{q_i}^2}{m_\phi^2}\right]
  \right)
  \mathcal{O}_{1,bh} + \frac14 \mathcal{O}_{2,bh} \right] ,
\end{align}
where the subscript $h$ refers to the decaying heavier flavor of DM.
The operators are given by
\begin{align}
  \mathcal{O}_{1,bh} &= 
  \left[ 
  \bar{\chi}_b 
  \gamma^\mu(1-\gamma^5)\partial^\nu 
  \chi_h\; 
  +h.c.\right] F_{\mu\nu} \,,
  \\ 
  \mathcal{O}_{2,bh} &= 
  \left[
  i\bar{\chi}_b 
  \gamma^\mu(1-\gamma^5)\partial^\nu 
  \chi_h\; 
  +h.c.\right]
  F^{\sigma\rho} \epsilon_{\mu\nu\sigma\rho} \,,
  \label{eq:ops}
\end{align}
and we have only kept the leading terms in momentum transfer,
$\mathcal{O}(k^2/m_\phi^2)$, and in $m_{q}/m_\phi$.

\begin{figure}[tp]
  \centering
  \includegraphics[width=0.28\textwidth]{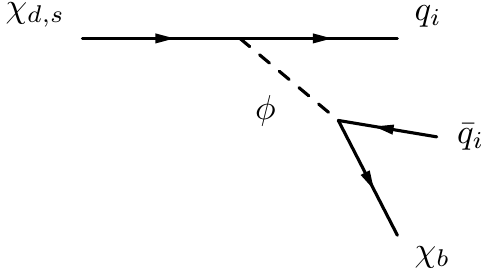}
  \qquad
  \includegraphics[width=0.28\textwidth]{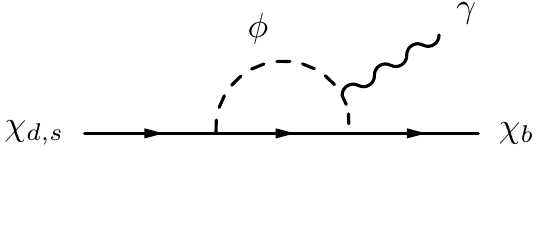}
  \qquad
  \includegraphics[width=0.28\textwidth]{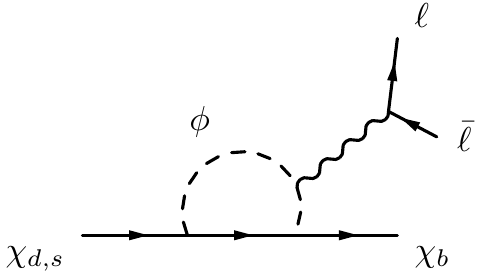}
  \caption{Decays of heavier flavors of DM into $\chi_b$.}
  \label{fig:decays}
\end{figure}

We see immediately that operator $\mathcal{O}_1$ does not give rise to
an on-shell photon decay, since the amplitude in that case is proportional to
$k^2$. Similarly, the operator $\mathcal{O}_2$ contribution is seen to
vanish because of a GIM-like cancellation. This is reminiscent of the
decay $b \to s\gamma$ in the SM, where an analogous GIM cancellation
causes the dominant decay to arise at a 
higher order in the EFT.
We can then estimate the decay rate by the SM partonic $b\to s\gamma$
rate. We replace $G_F$ by the coefficient of the effective operator
above, and account for an additional color factor. While 
the SM has spin-1 $W$ bosons running in the loop instead of
the scalar $\phi$, this estimate should suffice for our purposes.
The dominant contribution comes from the heaviest quark, i.e. the
$b$-quark in the loop. The decay rate estimate is 
\begin{align}
\nn
  \Gamma (\chi_h \to \chi_b \gamma)
  &\sim
  \frac{9
  |\lambda_{bh} \lambda_{bb}|^2 
  }{2\pi}
  \left[
  \frac{m_b^2}{m_\phi^2}
  \frac{e\  m_\chi
  }{
  (64 \pi^2 m_\phi^2)
  }
  \right]^2
  \,
  \delta m^3\,,
  \\&\sim
  |\lambda_{bh} \lambda_{bb}|^2 
  \left(
  \frac{1}{10^5\ \mathrm{ sec}}
  \right)
  \left(
  \frac{1000\ \mathrm{ GeV}}{m_\phi}
  \right)^8
  \left(
  \frac{m_\chi}{10\ \mathrm{ GeV}}
  \right)^2
  \left(
  \frac{\delta m}{0.2\ \mathrm{ GeV}}
  \right)^3 ,
\end{align}
where $\delta m$ is the mass splitting between the heavy and the
$b$-flavored DM. We see that this decay mode is too slow unless the dark
matter is relatively heavy and the splitting is sizeable.

We now estimate the third contribution in figure~\ref{fig:decays}. The
operator $\mathcal{O}_1$ in equation~\eqref{eq:ops} induces a
four-fermion contact operator of the DM with SM leptons and light
quarks.  We can hence calculate the three-body decay rate in analogy
with muon decay.  The presence of the log prevents the GIM-like
cancellation in this case.  The estimate for the decay rate to light
SM fermions is given as,
\begin{align}
  \Gamma (\chi_h \to \chi_b f \bar{f})
  &\sim
  |\lambda_{bh} \lambda_{bb}|^2 
  \left[
  \frac{e^2 }{96 \pi^2 m_\phi^2}
  \log\left[\frac{m_b^2}{m_{q,h}^2}\right]
  \right]^2
  \frac{9\  \delta m^5}{192 \pi^3}
  \,,
  \nn\\&\sim
  |\lambda_{bh} \lambda_{bb}|^2 
  \left(
  \frac{1}{1\ \mathrm{ sec}}
  \right)
  \left(
  \frac{1000\ \mathrm{GeV}}{m_\phi}
  \right)^4
  \left(
  \frac{\delta m}{0.2\ \mathrm{GeV}}
  \right)^5
  ,
\end{align}
where $m_{q,h}$ is the mass of the quark associated with the heavy
flavor.
We see that for relatively large values of the mass splitting,
the DM decays well before the BBN epoch. On the edges of our
parameter space, a more careful calculation is needed, especially in
the region of small mixing angles.

\end{appendix}


\bibliography{fdm}
\bibliographystyle{JHEP}

\end{document}